\documentclass[usenames,dvipsnames]{aa} 
\usepackage{natbib}

\usepackage{graphicx}
\usepackage{txfonts}
\usepackage{hyperref}
\usepackage{amsmath}
\usepackage{siunitx}
\usepackage{tablefootnote}
\usepackage[table]{xcolor}
\usepackage{xcolor}
\usepackage{amssymb}
\usepackage{pifont}
\usepackage{mathrsfs}  

\begin{document} 
    \title{Modelling Deuterated Isotopologues of Methanol toward the Pre-Stellar Core L1544}

   \author{W. Riedel \inst{1}, 
           O. Sipilä \inst{1},
           E. Redaelli \inst{1},
           P. Caselli \inst{1},
           A.I. Vasyunin \inst{2},
           F. Dulieu \inst{3}, and
           N. Watanabe \inst{4}
         }

   \institute{Max-Planck-Institut für extraterrestrische Physik, Gießenbachstraße 1, 85748 Garching bei
   	München, Germany\\
              \email{riedel@mpe.mpg.de}
         \and
             Ural Federal University, 620002, 19 Mira street, Yekaterinburg, Russia
        \and 
            CY Cergy Paris Université, Sorbonne Université, Observatoire de Paris, PSL University, CNRS, LERMA, F-95000 Cergy, France
        \and
            Institute of Low Temperature Science, Hokkaido University, Sapporo, Hokkaido, 060-0819, Japan
            }

   \date{Received -/ Accepted -}

 
  \abstract
  {In the extremely cold and dark environments of pre-stellar cores, methanol is formed on the surface of interstellar dust grains and released to the gas phase by non-thermal desorption mechanisms. Gaseous methanol constitutes the starting point for the formation of many massive complex organic molecules and is therefore of utmost importance for the build-up of chemical complexity.}
  {We aim to improve a previous model for the prediction of column densities and deuterium fractions of non- and singly deuterated methanol. Thereby, we try to identify crucial chemical and physical parameters, for which the study of deuteration could provide valuable additional constraints.}    
  {We employed a gas-grain chemical code to devise a model that is in agreement with the observed column density and deuterium fraction profiles of the innermost region of the pre-stellar core L1544. For that purpose, we developed a new treatment of reactive desorption, deriving an individual reactive desorption efficiency for every product species in a chemical reaction, that depends on the reaction enthalpy and type of underlying surface. Furthermore, we explored several options to promote the diffusion of hydrogen and deuterium atoms over the surface of interstellar dust grains, in order to increase methanol formation.}
  {Our fiducial model employs diffusion by quantum tunneling of hydrogen and deuterium atoms, resulting in CH$_3$OH and CH$_2$DOH column densities that are approximately an order of magnitude lower than the observed values, which improves the results compared to the previous model by a factor 10. The $N$(CH$_2$DOH)/$N$(CH$_3$OH) ratio is reproduced within a factor of 1.2 for the centre and 1.8 for the position of the methanol peak. Given the large uncertainties that chemical models typically have, we consider our predictions to be in agreement with the observations.  In general, we conclude that a diffusion process with a high diffusion rate needs to be employed to obtain methanol column densities that are in accordance with the observed values. Also, we find that the introduction of abstraction reactions into the methanol formation scheme suppresses deuteration, when used in combination with a high diffusion rate.
  } 
  {}

    
   \titlerunning{Modelling Deuterated Isotopologues of Methanol}
   \authorrunning{Riedel et al.}
   \maketitle
%

\section{Introduction} \label{ch:introduction}

Methanol (CH$_3$OH) is the simplest O-bearing complex organic molecule (COM) in the interstellar medium and an important precursor for saturated, more massive, COMs that are formed in the gas phase. Previous works, however, showed theoretically (\citealt{garrod2006}) and experimentally (\citealt{geppert2005}) that the presumed gas phase formation route is very inefficient and unable to account for the observed gas phase abundances. Simultaneously, the formation of methanol on the surface of dust grains was proposed and investigated in an extensive manner experimentally in several independent projects (e.g.: \citealt{watanabe2002}, \citealt{fuchs2009}). It was concluded that formaldehyde and methanol can be produced by multiple successive addition reactions of CO with diffusive hydrogen atoms. The measured significant isotope effect for hydrogen and deuterium suggests that hydrogenation and deuteration on the surface proceeds via quantum tunneling reactions (\citealt{hidaka2007}) and is therefore progressing with a much higher rate than in the gas phase. Additionally, the existence of so-called abstraction reactions, that remove H$_2$ from the molecule and thereby reverse the addition reaction, was postulated and found in various laboratory experiments (e.g.: \citealt{hidaka2009}, \citealt{minissale2016c}). However, there seems to be some disagreement about the exact reaction scheme and the magnitude of the reaction rates between the different experimental approaches. 

For the production of more advanced COMs in the gas phase, methanol needs to desorb from the surface of the dust grains. In hot cores and corinos, the desorption of the molecular contents of the surface phase does proceed very efficiently by thermal evaporation and photoevaporation. In the cold and dark environment of pre-stellar cores, however, these mechanisms are negligible. Therefore, only very low abundances of methanol were expected to exist in the gas phase of pre-stellar cores. Surprisingly, several surveys, e.g.: \cite{bacmann2012}, \cite{cernicharo2012} and \cite{jimenez2016}, conducted toward dark molecular clouds, found comparatively high abundances of methanol and other complex organic molecules. 

In order to explain these unexpected findings, astrochemical models need to employ other mechanisms for the evaporation of surface molecules. One promising candidate, reactive desorption, is based on the energy released in an exothermic chemical reaction, which can lead to the desorption of the reaction product(s). Nowadays, most chemical codes include a simple treatment of reactive desorption following the recipe described by \cite{garrod2007}. According to those authors, a (nearly) constant reactive desorption efficiency, of typically 1\%, is employed independent of the desorbing molecule. However, recent laboratory experiments found a strong dependence of the reactive desorption efficiency on the chemical reaction and the type of underlying surface (\citealt{minissale2016b, chuang2018}). 

In this paper, we develop an updated version of the description of reactive desorption presented by \cite{vasyunin2017}, itself based on the experiments by \cite{minissale2016b}. The motivation for our work are the column density maps of CH$_3$OH and CH$_2$DOH and theoretical predictions presented by \cite{chacon(2019)}. They carried out single-dish observations with the IRAM 30m telescope (also \citealt{bizzocchi2014, vastel2014, spezzano2016}), showing that CH$_3$OH peaks in an asymmetric ring around the dust peak with the strongest emission in the northern part of the pre-stellar core. \cite{chacon(2019)} analysed the formaldehyde and methanol column densities along a cut set by the position of the dust and the offset methanol peak. 

An important aim of this work is to improve the theoretical column density profiles (hereafter model S16) made by the chemical code $\textit{pyRate}$ (\citealt{sipilae2015a, sipilae2019b}), particularly to get more accurate predictions about the deuteration of methanol. Additionally, we compare our predictions with the results of the chemical code presented in \cite{vasyunin2017} (hereafter V17 model). These two models were used in the work of \cite{chacon(2019)}, where it was found that the V17 model
produced acceptable results for non-deuterated methanol (CH$_3$OH). However, the column density profile for singly deuterated methanol (CH$_2$DOH) had to be obtained by using the N(CH$_2$DOH)/N(CH$_3$OH) ratio predicted by $\textit{pyRate}$, as the MONACO code (\citealt{vasyunin2017}) does not treat deuteration, whereas  $\textit{pyRate}$ includes an extensive description of deuterium chemistry.  To carry out a similar analysis in a consistent manner, we implement here an updated version of the reactive desorption mechanism used in \cite{vasyunin2017} into \textit{pyRate}.

\cite{vasyunin2017} and \cite{chacon(2019)} both found that diffusion by quantum tunneling of atomic and molecular hydrogen had to be employed in order to explain the observed methanol column density profile. In the present paper, we employ tunneling diffusion for hydrogen and deuterium atoms for our fiducial model, neglecting the tunneling diffusion of molecular hydrogen. However, we also explore the option to promote thermal diffusion by decreasing the diffusion-to-binding energy ratio $E_{\mathrm{d}}/E_{\mathrm{b}}$, which determines the threshold for the diffusion of molecules over the surface of an interstellar dust grain via thermal hopping, from 0.55 to the lowest value reported in the experimental literature (0.2; \citealt{furuya2022}). Moreover, we test several alternative sets of input parameter and how well they are able to explain the observed deuterium fraction profile.  

The paper is structured as follows: Section \ref{ch:model} describes the new reactive desorption mechanism and the chemical and physical model in detail. Section \ref{ch:results} describes the results of the fiducial model and compares it with the observationally obtained column density and deuterium fraction profiles by \cite{chacon(2019)} and with the model presented in \cite{vasyunin2017}. Section \ref{ch:discussion} discusses multiple modifications to the chemical and physical parameters and their effects on the results. Section \ref{ch:conclusion} presents our conclusions. Appendices \ref{ch:App_A} and \ref{ch:App_B} provide additional information on the used chemical parameters and reaction schemes. 

\begin{figure*} 
	\includegraphics*[width=1.0\textwidth]{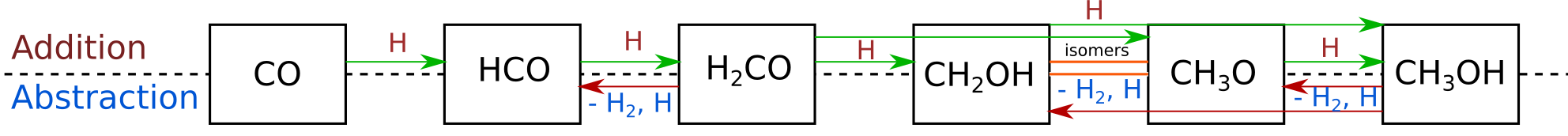}
	\caption{Reaction scheme for the formation of non-deuterated CH$_3$OH by successive hydrogenation. Hydrogen molecules can also be segregated from CH$_3$OH or its precursors by abstraction reactions.} 
	\label{fig:methanol_formation}
\end{figure*}

\section{Model} \label{ch:model}
\subsection{Treatment of Reactive Desorption} \label{ch:reactive_desorption}
For the extension of the reactive desorption mechanism, we adopt the physical scenario proposed and experimentally justified by \cite{minissale2016b}. The mechanism is implemented following \cite{vasyunin2017}, but it is modified for this paper, as detailed below. 

\cite{minissale2016b} developed a formula, that expresses the reactive desorption efficiency $RD$ as a function of reaction and surface-dependent properties:  
\begin{align} \label{eq:RD}
&RD = \exp\left(- \frac{E_{\mathrm{b}} N}{\epsilon \Delta H}\right) ,
\end{align} 
where $\Delta H$ is the reaction enthalpy, $E_{\mathrm{b}}$ the binding energy, $N$ the number of the degrees of freedom (translational, rotational and vibrational) of the reaction product and $\epsilon$ the fraction of kinetic energy retained by the reaction product. We assume that the available reaction enthalpy is distributed equally into all the degrees of freedom $N$ and that only the energy going into the vertical translational degree of freedom is used for the desorption of an atom or a molecule from the surface of a dust grain. Therefore, for one-product reactions, 1/$N$ of the reaction enthalpy is distributed into motion perpendicular to the surface of the dust grain. The number of degrees of freedom $N$ of the product  can be simply derived by $N = 3n_{\mathrm{atoms}}$ with $n_{\mathrm{atoms}}$ being the number of atoms of the reaction product. 
This approach has the effect that the less complex molecules, consisting of fewer atoms, can use a larger share of the reaction enthalpy for vertical motion off the surface. For the more complex molecules, the total number of degrees of freedom increases quickly with increasing number of atoms, due to the fact that larger molecules have more possibilities for vibrational excitation. However, the number of translational degrees of freedom stays the same. As a consequence, a smaller share of the reaction enthalpy is converted into vertical motion. For example, a forming CO molecule, with only 2 atoms, receives 1/6 of the reaction enthalpy for vertical motion, while a forming CH$_3$OH molecule, with 6 atoms, receives only 1/18.

The dependency of the reactive desorption efficiency on the type of surface is expressed by the fraction of kinetic energy $\epsilon$ received by the reaction product with mass $m$ when colliding with a surface element with an effective mass $M$:
\begin{align} \label{eq:kinetic_energy}
\epsilon = \left(\frac{M - m}{M+ m}\right)^2 .
\end{align}
In the experiments by \cite{minissale2016b}, the effective mass $M$ is a parameter that needs to be fitted. It was found to be typically much larger than the mass of a single atom or molecule of the surface species, but is more consistent with a collective behaviour of also the neighbouring molecules that is induced by the rigidity of the surface. The more rigid the surface, the higher the effective mass of the surface element $M$ and the easier it is for the reaction products to bounce off from the surface into the gas phase. 
For the effective masses we have mostly kept the values adopted by \cite{vasyunin2017}, namely $M=\SI{48}{amu}$ for a H$_2$O surface and $M=\SI{100}{amu}$ for a CO-surface, with the exception of the effective mass for bare grain, where we have taken a somewhat lower value of $M = \SI{120}{amu}$ that has been suggested as a common value for carbonaceous and silicate grains following the experiments conducted by \cite{minissale2016b}. Every surface which does not consist of H$_2$O or bare grain is treated as if it were CO. Considering that grain surfaces in the inner, very cold regions of pre-stellar cores are covered, aside from CO itself, by species with a similar molecular weight (e.g.: N$_2$, H$_2$CO and CH$_3$OH), this seems to be a reasonable approximation given the lack of more detailed measurements.
To consider that a grain surface might be covered by a mixture of bare surface, H$_2$O or CO, we follow the average surface composition of the dust grains over time, starting with a bare grain surface, which is then quickly covered by water ice and later on CO. Then, we scale the individual reactive desorption efficiencies for a reaction $i$ by the fraction of the surface sites that are covered by the particular surface type~$j$: 
\begin{align}
RD_{\mathrm{tot}}(i,t) = \sum_j RD_{\mathrm{j}}(i) \cdot \frac{n_j^*(t)}{n_{\mathrm{tot}}^*(t)} ,
\end{align}
where $RD_{\mathrm{j}}$ is the individual reactive desorption efficiency for either bare grain, H$_2$O or CO, $n_j^*$ the number of surface sites inhabited by surface type $j$ and $n_{\mathrm{tot}}^*$ the total number of surface sites.                        
The largest difference between the procedure applied in this paper and the one presented in \cite{vasyunin2017} is the treatment of reactions with more than one reaction product. \cite{vasyunin2017} treated multi-product reactions identically to one-product reactions, so that every reaction product receives the entire available reaction enthalpy. This approach clearly violates energy conservation. However, this simplification was considered to be negligible for the chemical network that was used in \cite{vasyunin2017} as it includes mostly two-product reactions involving heavier reactants that proceed inefficiently at the low grain temperatures typical of pre-stellar cores.

In contrast, the method presented in this paper includes a recipe for the mass-dependent partitioning of the total reaction enthalpy in the case of a two-product reaction. We assume that the desorbing molecules are first isolated from the surface and then undergo an elastic collision with the surface of the grain to gain velocity in the vertical direction. Considering the conservation of linear momentum, one can derive an expression for the kinetic energy $E_{\mathrm{kin}}$ received by a product species $i$:  

\begin{align} \label{eq:kinetic_energy}
E_{\mathrm{kin}}(\mathrm{i}) = E_{\mathrm{kin}} \cdot \frac{m_{\mathrm{j}}}{m_{\mathrm{i}} + m_{\mathrm{j}}} = \frac{N_{\mathrm{trans}}}{N_\mathrm{tot}} \cdot \Delta H \cdot \frac{m_{\mathrm{j}}}{m_{\mathrm{i}} + m_{\mathrm{j}}}   .
\end{align}

The kinetic energy received by species i is scaled by the fraction of the mass of the other reaction partner (species j) and the combined mass of both reaction products.
Additionally, we derive the share of the reaction enthalpy that goes into kinetic energy as the fraction of the number of translational degrees of freedom to the total number of degrees of freedom. Note that the number of translational degrees of freedom is the combined number for both reactants: $N_{\mathrm{trans}} = 3n_{\mathrm{prod}}$, where $n_{\mathrm{prod}}$ is the number of products. The total number of degrees of freedom is defined similarly as the sum of the individual degrees of freedom for both reaction partners: 
$N_{\mathrm{tot}}$ = $\sum_i 3n_{\mathrm{atoms,i}}$, with $n_{\mathrm{atoms,i}}$ being the number of atoms of reaction product i.
Here, we consider that the vertical motion that is responsible for the desorption is only one of three translational degrees of freedom. This introduces an additional factor of 1/3.
Finally, we can derive an equation for the reactive desorption efficiency of a two-product reaction (\ref{eq:RD_2products}), that simplifies to Eq. (\ref{eq:RD}) in the case that the occurring reaction has only one reaction product:
\begin{align} \label{eq:RD_2products}
RD_{\mathrm{i}} = \exp\left( -\frac{1}{\epsilon} \left( \frac{E_{\mathrm{b}}}{\Delta H} \right) \left(\frac{1}{3} \frac{N_{\mathrm{trans}}}{N_\mathrm{tot}} \frac{m_\mathrm{j}}{m_{\mathrm{i}} + m_{\mathrm{j}}} \right)^{-1} \right) .
\end{align}

The new approach has the effect that the lighter of the two products gets the larger share of the kinetic energy and consequently has an increased reactive desorption efficiency, while the one for the heavier product is decreased. For example, considering the reaction:
\begin{align}
    \text{H}_2\text{CO} + \text{H} \rightarrow \text{HCO} + \text{H}_2 ,
\end{align}
with HCO being approximately 14 times more massive than H$_2$, the HCO molecule receives only 1/15 of the kinetic energy, whereas the molecular hydrogen gets 14/15. This results in a reactive desorption efficiency for a CO surface of 54.0\% for the H$_2$ molecule and only \num{1.5e-36}\% for the HCO molecule. The procedure presented in \cite{vasyunin2017} would provide an efficiency of 63.1\% for H$_2$ and 0.1\% for the HCO molecule. However, there are no similar reactions efficient at low temperatures in the \cite{vasyunin2017} model.

\subsection{Chemical Model} \label{ch:chemical_model}

We incorporate the new description of reactive desorption into the rate-equation based chemical code $\textit{pyRate}$, described in more detail in \cite{sipilae2015a, sipilae2019b}. It tracks the chemical evolution both in the gas phase and on the grain surface. The basis of the chemical network is the 2014 public release of the KIDA gas phase network (kida.uva.2014, \citealt{wakelam2015}), which was extended by deuterium chemistry for molecules with up to seven atoms. The code also tracks the various spin states of the light hydrogen-bearing species H$_2$, H$_2^+$ and H$_3^+$ and their deuterated isotopologues, as well as multiply protonated or deuterated species involved in  the water and ammonia formation networks. All together, the network includes $\approx$ 74000 gas phase reactions and $\approx$ 2100 grain surface reactions. For the inclusion of abstraction reactions into the methanol formation pathway in the models presented in this work, we refer to the chemical network proposed by \cite{hidaka2009} and depicted in Figure \ref{fig:methanol_formation} (representation with H) and Figure \ref{fig:network_EA} (more extensive representation with H and D). We adopt atomic initial abundances (see Table \ref{tab:initial_abundances}) taken from \cite{semenov2010}, as they were also used for the S16 model, whose improvement is the main aim of this work.
Also, we employ a three-phase model, consisting of a gas phase, a chemically-active surface phase and a chemically-inert mantle phase. The dust grains are assumed to be spherically symmetric with a radius of \SI{0.1}{\micro\meter}.

For exploring our new treatment of reactive desorption, the choice of binding energies $E_{\mathrm{b}}$ and formation enthalpies $H_{\mathrm{form}}$ is crucial. The values of $E_{\mathrm{b}}$ and $H_{\mathrm{form}}$ are displayed in Table (\ref{tab:formation_enthalpies}). The binding energies are taken from \cite{semenov2010}. Most of the formation enthalpies are adopted from \cite{du2012}. For the other remaining species, the data sources are marked in Table \ref{tab:formation_enthalpies} in the appendix. Unfortunately, experimental values for deuterated molecules are quite scarce. For this reason, we apply the same values as for the non-deuterated isotopologues, with the exception of the species marked with a star in Table (\ref{tab:formation_enthalpies}), for which we found individual values in the NIST Chemistry WebBook \footnote{https://webbook.nist.gov/chemistry}.

\begin{table}
	\caption{Initial chemical abundances with respect to n$_{\mathrm{H}}$.}
	\label{tab:initial_abundances}
	\begin{tabular}{l l l l}
	\hline
	\hline
	\noalign{\smallskip}
	 \text{Species}     &  \text{Initial abundance} & \text{Species} & \text{Initial abundance}\\
	\noalign{\smallskip}
	\hline
	\noalign{\smallskip}
	$\mathrm{He}$ & $9.00 \times 10^{-2}$ & $\mathrm{S}^+$ & $8.00 \times 10^{-8}$   \\
	$\mathrm{pH}_2$ & $5.00 \times 10^{-1}$ & $\mathrm{Si}^+$ & $8.00 \times 10^{-9}$   \\
	$\mathrm{oH}_2$ & $5.00 \times 10^{-4}$ & $\mathrm{Na}^+$ & $2.00 \times 10^{-9}$ \\
	$\mathrm{HD}$ & $1.60 \times 10^{-5}$ & $\mathrm{Mg}^+$ & $7.00 \times 10^{-9}$ \\
	$\mathrm{H}$ &  $1.00 \times 10^{-8}$ & $\mathrm{Fe}^+$ & $3.00 \times 10^{-9}$  \\
	$\mathrm{D}$ & $1.00 \times 10^{-8}$ &  $\mathrm{P}^+$ & $2.00 \times 10^{-10}$  \\
	$\mathrm{C}^+$ & $1.20 \times 10^{-4}$ & $\mathrm{Cl}^+$ & $1.00 \times 10^{-9}$ \\
	$\mathrm{N}$ & $7.60 \times 10^{-5}$ & $\mathrm{F}$ & $2.00 \times 10^{-9}$ \\
	$\mathrm{O}$ & $2.56 \times 10^{-4}$ &  \\
	\noalign{\smallskip} 
	\hline
	\end{tabular}
 \tablefoot{\tablefoottext{a}{from \cite{semenov2010}}}
\end{table}

\subsection{Physical Model} \label{ch:physical_model}
\begin{figure*} 
	\includegraphics*[width=1.0\textwidth]{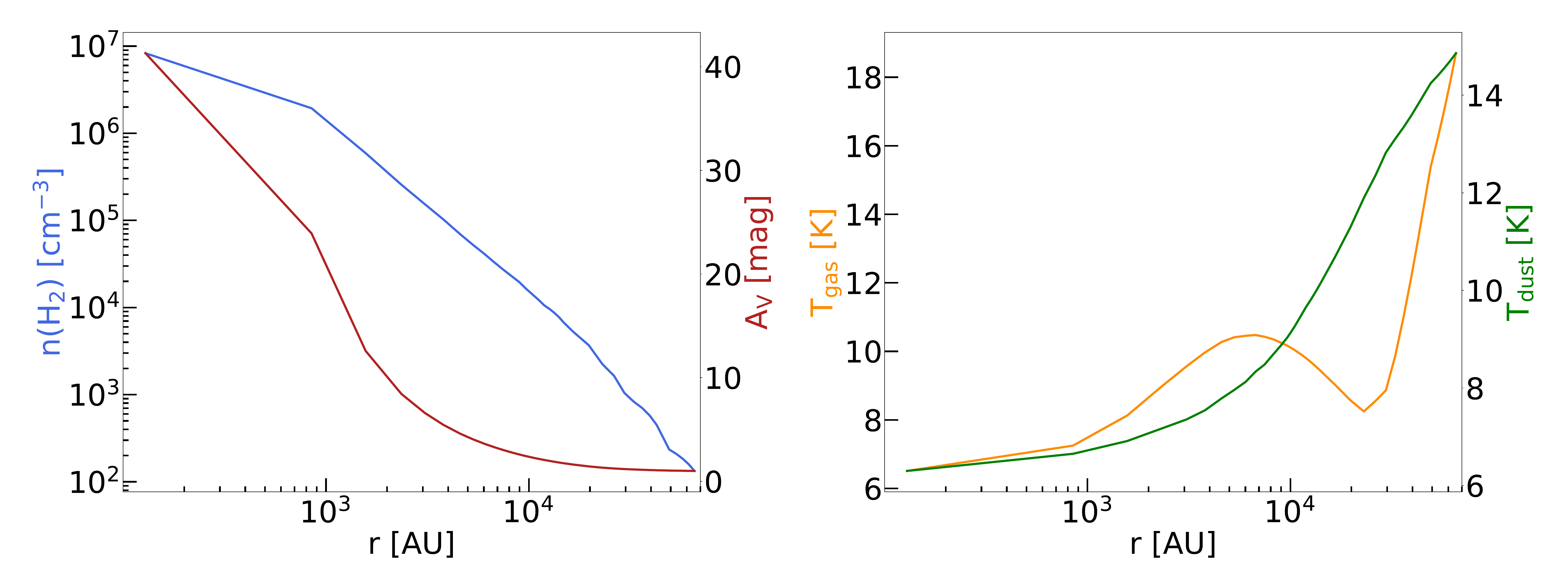}
	\caption{Physical model developed in \cite{keto&caselli2010} yielding static radial profiles of the H$_2$ density $n(\mathrm{H_2})$ (blue, in logarithmic scale), the visual extinction $A_{\mathrm{V}}$ (red), the gas temperature $T_{\mathrm{gas}}$ (orange) and the dust temperature $T_{\mathrm{dust}}$ (green).}
	\label{fig:physical_parameter}
\end{figure*}

As mentioned in Section \ref{ch:introduction}, one of the main aims of this work is to improve the theoretical predictions made by the chemical code $\textit{pyRate}$ for the column density profiles of CH$_3$OH and CH$_2$DOH and to compare these again to the observational (\citealt{chacon(2019)} and theoretical profiles from the V17 model (\citealt{vasyunin2017}). For that purpose, we extensively explored the chemical evolution in the pre-stellar core L1544 with a one-dimensional physical model. It was derived from the one presented in \cite{keto&caselli2010} and described in more detail in \cite{sipilae2019a}. The model provides radius-dependent, but time-independent, values for the H$_2$-density $n(\mathrm{H}_2$), the gas temperature $T_{\mathrm{gas}}$, the dust temperature $T_{\mathrm{dust}}$ and the visual extinction $A_{\mathrm{V}}$ as shown in Figure \ref{fig:physical_parameter}. The core model consists of 35 concentric shells spanning the core radius of $\SI{0.32}{pc}$. The chemistry is solved separately for each shell, yielding a spherically-symmetric spatio-temporal evolution of molecular abundances. We calculate column densities by integrating along the line-of-sight for different impact factors from the core centre. Afterwards, the column density distribution is convolved with a 30'' gaussian beam, corresponding to the angular resolution of the observations by \cite{chacon(2019)}.

Taking the core model as a basis, we have run several simulations varying multiple chemical and physical parameters. All presented models include the new treatment for reactive desorption with individual efficiencies for every exothermic surface reaction and various surface types, as described in Section \ref{ch:reactive_desorption}. An overview of the various models is presented in Table \ref{tab:overview_models}.

\section{Results} \label{ch:results}

\subsection{Fiducial model} \label{ch:fiducial_model}
We selected the 1D-4 model as our fiducial model, because it is the closest to V17 in terms of parameter space.
In the fiducial model, the diffusion of H and D atoms by quantum tunneling is enabled, while, abstraction reactions as shown in Figure \ref{fig:methanol_formation}  (or their deuterated analogues) are not included. This choice is discussed in more detail in Section \ref{ch:abstraction_diffusion}. 

\begin{figure*} 
    \centering
	\includegraphics*[width=1.0\textwidth]{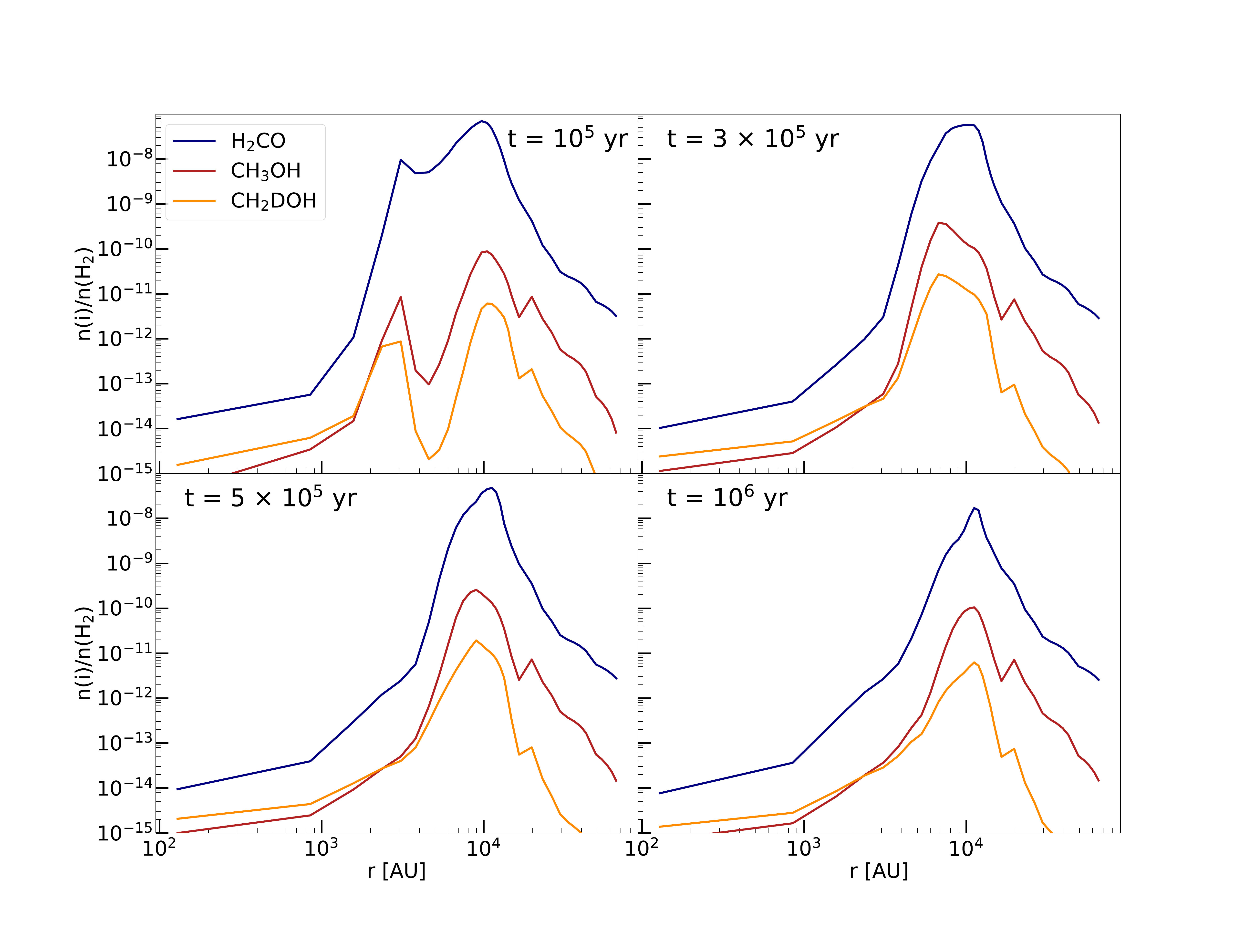}
	\caption{Gas phase abundance profiles of H$_2$CO, CH$_3$OH and CH$_2$DOH for the 1D-4 model for four different time steps ranging from 10$^5$ yr (left) to 10$^6$ yr (right). The best fit time is  $t$ = \SI{3e+5}{yr}.}
	\label{fig:abundance_profiles_fid_model}
\end{figure*}

We calculated column density profiles for all species observed by \cite{chacon(2019)}, for several time steps in the range of $\SI{e5}{yr}$ to $\SI{e6}{yr}$ in the fiducial model. Figure \ref{fig:abundance_profiles_fid_model} shows the molecular abundances with respect to H$_2$ and Figure \ref{fig:column_density_profiles_fid_model} shows the column density profiles for H$_2$CO, CH$_3$OH and CH$_2$DOH for four different time steps (\SI{1e5}{yr}, \SI{3e+5}{yr}, \SI{5e5}{yr} and \SI{1e6}{yr}). Note that these species freeze out onto dust grains in the very cold centre of the pre-stellar core and then peak at a density of n(H$_2$) $\approx$ 10$^4$ cm$^{-3}$, which corresponds to a radius of $\approx \SI{5200}{AU}$ in the theoretical profiles, offset from the position of the dust peak. For a time of $t = \SI{3.5e+5}{yr}$, the CH$_3$OH abundance reaches its maximum value of approximately $\num{4e-10}$, which is in good agreement with values observed in various pre-stellar cores (e.g.: \citealt{scibelli&shirley2010}, \citealt{harju2020}, \citealt{spezzano2020}, \citealt{punanova2022}).

In \cite{chacon(2019)}, the time step in the S16 model was chosen such that the simulated CO column density matched approximately the observed one (\citealt{caselli1999}), which occurred at a very early time of \SI{3e4}{yr}. However, S16 did not include CO self-shielding, and is therefore probably underestimating the total amount of CO. For this very early time step, the S16 model produces a centrally flat CH$_3$OH column density profile with an amplitude of $N \approx$ \SI{1e11}{\per\square\centi\meter}, under-producing the observed values for methanol of $\SI{3.9e13}{\per\square\centi\meter}$ at the dust peak and $\SI{5.9e13}{\per\square\centi\meter}$ at the methanol peak by roughly two orders of magnitude. On the other hand, the $N$(CH$_2$DOH)/$N$(CH$_3$OH) ratio in the S16 model approaches unity, thereby overestimating the observed deuterium fraction of \num{0.07} by a factor of 10. We note that the chemical network that was used for the S16 model includes reactions like CH$_2$DOH + H$_3$O$^+$ $\rightarrow$ CH$_3$OHD$^+$ + H$_2$O, or analogues for other deuterated forms of methanol, with the effect of increasing the deuterium fraction. We do not permit this sort of exchange to happen in the revised chemical network, as this would require the addition of a (hydrogen) atom as well as the exchange of a hydrogen and a deuterium atom between the different functional groups of methanol, which we consider to have a low probability.

\begin{figure}
	\includegraphics*[width=0.5\textwidth]{column_density_profiles_fid_model.pdf}
	\caption{Column density profiles of H$_2$CO, CH$_3$OH and CH$_2$DOH for the 1D-4 Model for four different time steps ranging from 10$^{5}$ yr to 10$^6$ yr. The lines with markers show the modelled results, integrated along the line of sight and convolved with a 30'' beam. The solid lines show the observationally obtained column density profiles obtained by \cite{chacon(2019)}, by taking a cut through the dust and methanol peaks. The grey-shaded areas indicate the error bars of the column densities. The position of the dust peak is at $r$ = \SI{0}{AU}, while $r$ > \SI{0}{AU} is the direction towards the methanol peak.}
	\label{fig:column_density_profiles_fid_model}
\end{figure}

To estimate a new best-fit time for the fiducial model, we calculated the $\chi^2$-values of the observed central column density vs. the corresponding values in the fiducial model for different time steps and species. Subsequently the time step where the sum of the $\chi^2$-values of all species is minimised was taken as the best-fit time. For the 1D-4 model, this corresponds to a best-fit time of $t = \SI{3e5}{yr}$, which is roughly consistent with estimations from other molecules (e.g. \citealt{redaelli2019,redaelli2021}). Additionally, it coincides with the occurrence of the highest methanol column density in the probed time frame. For $t = \SI{3.0e5}{yr}$, the fiducial model reaches a column density of methanol of $\approx$ \SI{1.5e+12}{\per\square\centi\meter}, which is roughly an order of magnitude lower than the column densities determined by the observations. Given the large uncertainties that chemical modelling typically experiences, this is an acceptable agreement (\citealt{vasyunin2004, vasyunin2008}, \citealt{wakelam2010} and references therein).

\begin{figure*} 
	\includegraphics*[width=1.0\textwidth]{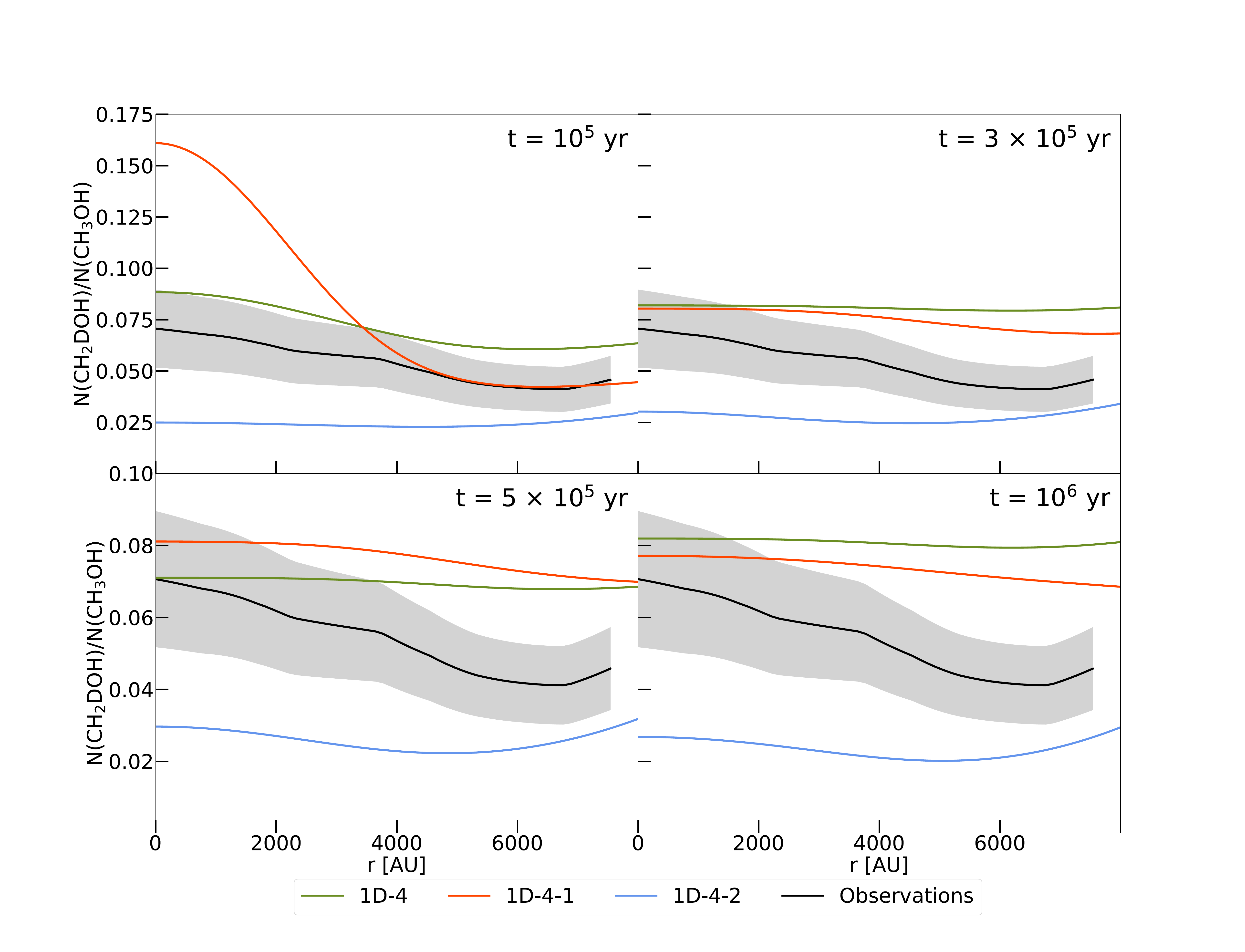}
	\caption{Modelled ratio between singly-deuterated methanol (CH$_2$DOH) and non-deuterated methanol (CH$_3$OH) for the 1D-4 Model for four different time steps ranging from 10$^{5}$ yr (top left) to 10$^6$ yr (bottom right). Additionally, we show two variations of this model: one with four layers in the  chemically active surface phase instead of one (1D-4-1), and one with a two-phase model (1D-4-2) (see also Table \ref{tab:overview_models} and Section \ref{ch:comparison_V17}) for a more detailed explanation.)The coloured lines shows the column density ratio of the models, while the black line indicates the observed ratio (errors as grey-shaded areas).}
	\label{fig:CH2DOH_CH3OH_fid_model}
\end{figure*}

Figure \ref{fig:CH2DOH_CH3OH_fid_model} compares the modelled deuterium fraction $N$(CH$_2$DOH)/$N$(CH$_3$OH) with the observed one. At the early time step ($t$ = \SI{1e+5}{yr}), the deuterium fraction profile is slightly higher than the observed one, but the shape of the profiles are nearly identical. The modelled deuterium fraction flattens with time. This behaviour can probably be explained by the use of the static physical model that is employed here. At the best fit time of $t$ = \SI{3e+5}{yr}, the deuterium fraction assumes an almost completely flat profile with a value of $\approx$ \num{0.08}. At the intermediate ($t$ = $\SI{5e5}{yr}$) and late ($t = \SI{1e6}{yr}$) time steps, the $N$(CH$_2$DOH)/$N$(CH$_3$OH) ratio starts to decrease slightly, but stays at a level that is in agreement with the observed one.

\subsection{Comparison with V17} \label{ch:comparison_V17}

\begin{figure*} 
	\includegraphics*[width=1.0\textwidth]{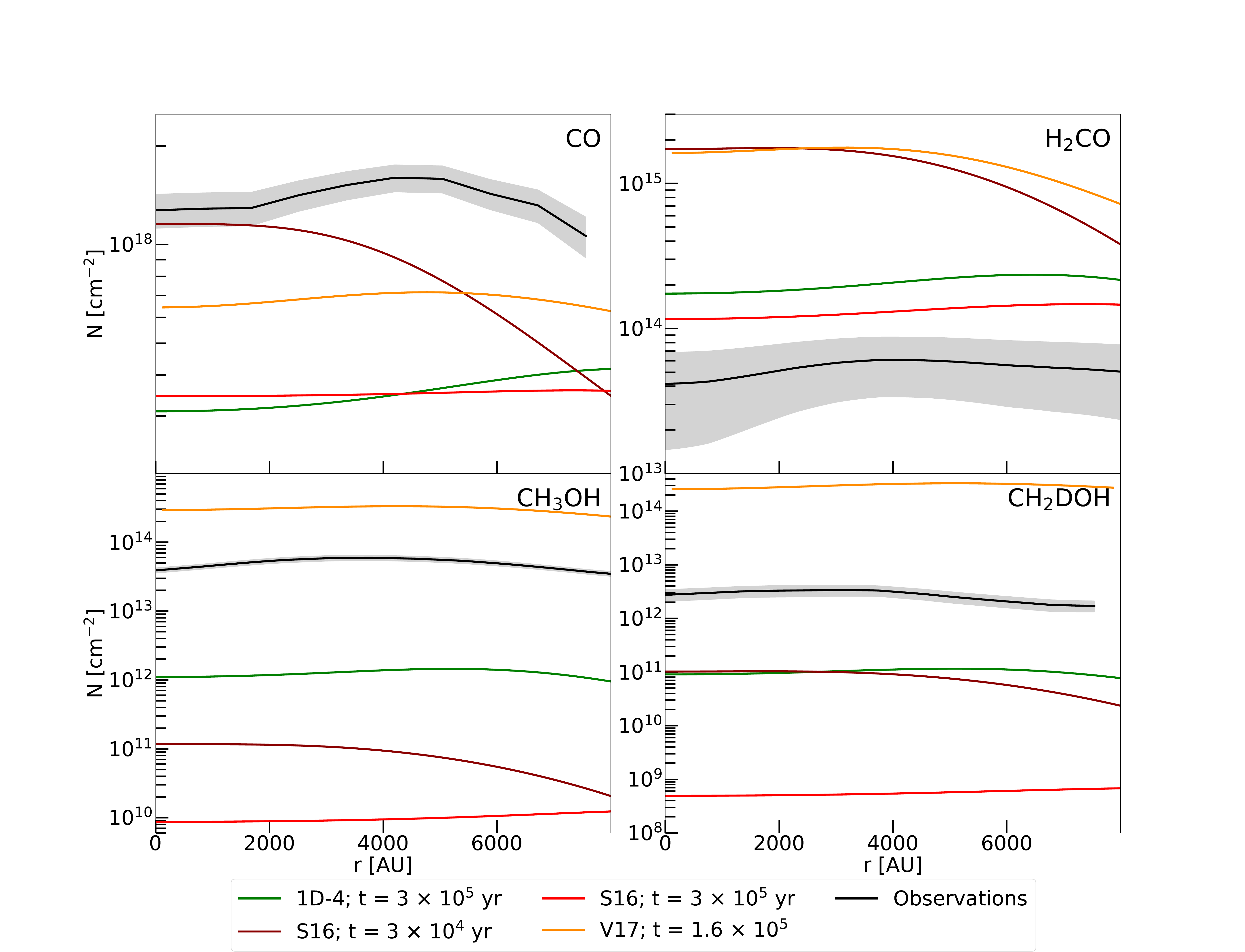}
	\caption{Comparison of the models presented in \cite{chacon(2019)} - S16 (obtained with \textit{pyRate}) and V17 (obtained with \textit{MONACO}) - to the 1D-4 model (fiducial model) computed with an updated version of \textit{pyRate}. The column density profiles are shown at different time steps: $t$ = \SI{1.6e+5}{yr} for V17, $t$ = \SI{3.0e+4}{yr} and $t$ = \SI{3.0e+5}{yr} for S16 and $t$ = \SI{3.0e+5}{yr} for 1D-4. The observed profiles are depicted in black (errors as grey-shaded areas). The CO column densities were obtained by using observations of C$^{17}$O and the average isotopic ratios of $^{16}$O/$^{18}$O = 557 and $^{18}$O/$^{17}$O = 3.6 in the local ISM (\cite{wilson1999}).}
	\label{fig:comparison_V17}
\end{figure*}

The fiducial model (1D-4) shows a much better agreement for the column densities of non-deuterated methanol than the S16 model, which was produced by an earlier version of $\textit{pyRate}$, when compared to the V17 model produced by the chemical code \textit{MONACO} and presented in \cite{vasyunin2017} and \cite{chacon(2019)} (see Figure \ref{fig:comparison_V17}). We have set the values of the various physical parameters to correspond to those in V17 as closely as possible. Still, several differences remain between $\textit{MONACO}$ and $\textit{pyRate}$, and these cannot be fully understood without a detailed direct comparison of the two models, which is out of the scope of this paper. 

Here, we point out some of the most noticeable differences.
The chemical models are not identical, as we are using different chemical networks. $\textit{MONACO}$ is specialised to describe the formation of complex organic molecules, while \textit{pyRate} concentrates on the description of deuteration. V17 does not consider deuteration at all. Also, \textit{pyRate}'s chemical network contains "backward" abstraction reactions adopted from \cite{hidaka2009}. We removed them for several models presented in this paper, including the fiducial model 1D-4, in order to compare our model to the V17 model. In V17, these reactions are deliberately left out due to their badly constrained reaction rates. 

Both codes use a so-called three-phase grain model with multiple layers as opposed to a simpler two-phase model. Three-phase models usually consist of three distinct phases: the gas, the ice surface and the bulk phase, which again can be subdivided into individual layers. Two-phase models only distinguish between gas and surface phase. In most astrochemical codes, the bulk is simply a chemically inert storage of accreted molecules, while the chemical reactions occur solely in the gas and surface phase. A two-phase model has no such storage. Every molecule on the grain can react or desorb at any time. The introduction of layers into a grain model enables the storage of molecules into the bulk in the order that they are accreted. Gas phase molecules can only accrete to the surface phase, where they can react with another surface molecule. The surface phase usually consists of only one layer. In the case that all binding sites in the surface phase are filled, molecules are transferred continuously into the bulk phase, which keeps growing.

The \textit{MONACO} code has a more advanced description of the physical processes taking place on dust grains than most other astrochemical codes. In \textit{MONACO}, the bulk experiences a slow type of diffusion and species are therefore able to meet one another and react, instead of being just stored away. Moreover, the surface phase consists not only of the uppermost layer, but of the first four layers in order to also allow atoms to be diffused in vertical direction. The fiducial model presented in this paper also uses a multilayer dust model, but with a chemically inactive bulk phase and only one layer in the surface phase, which reduces the surface area on which methanol and its precursors can be hydrogenated. 

To evaluate the consequences of our choice to use only one layer in the chemically active surface phase in our models, we have also tested a modification of our fiducial model, 1D-4-1, where we set the number of layers in the surface phase to four and, as a reference, a two-phase model of the fiducial model set up, 1D-4-2, where the entire ice is available for reactions. For 1D-4-1, we find that both the CH$_3$OH and the CH$_2$DOH column densities are increased by a factor of a few in comparison to the fiducial model in the entire considered time frame. This makes sense as with the increase of layers in the surface phase the model approaches the two-phase model, showing consistently higher column densities for both isotopologues of methanol. This behaviour is credited both to an increased overall production of methanol, but also an increased reactive desorption efficiency in the very centre of the pre-stellar core, due to a higher coverage of the surface with CO. However, the deuterium fraction of the two-phase model is quite low (see also Section \ref{ch:grain_model}), as the increase in column densities for CH$_2$DOH is lower than for CH$_3$OH. Surprisingly, this is not the case for the model with four layers in the surface phase, as is depicted in Figure \ref{fig:CH2DOH_CH3OH_fid_model}. Instead, we find a considerable increase of the deuterium fraction compared to the fiducial model for early time steps, reaching a value of up to $\approx$ \num{0.16} at $t$ = \SI{1e+5}{yr}. The deuterium fraction of the model with four layers in the surface phase drops slightly below that of the fiducial model for time steps around when the methanol column density peaks, but increases again above it for later time steps.

Both models, the fiducial and V17, employ a similar form of diffusion. The diffusion-to-binding energy $E_{\mathrm{d}}$/$E_{\mathrm{b}}$ is set to a constant value of \num{0.55}, as suggested by \cite{minissale2016a}, for all surface species. Additionally, both codes consider the inclusion of diffusion by quantum tunneling. Although, while $\textit{pyRate}$ only follows the tunneling diffusion of H (and D), $\textit{MONACO}$ also traces that of H$_2$.   

The mechanism for reactive desorption in the two codes is similar: altered here for chemical reactions with multiple products, to ensure that it adheres to conservation of energy, and identical for reactions with a single product. If one does not introduce abstraction reactions in the formation scheme of methanol, reactions with two reaction products are considered to be negligible, as the formation just proceeds by successive addition reactions of H. Note that even though the procedure for single product reactions is the same in both codes, they can produce unequal reactive desorption efficiencies owing to the fact that we use different sets of enthalpies and binding energies. The binding energies and formation enthalpies for the models presented in this work and in the S16 model are adopted from \cite{semenov2010}, \cite{du2012}, the NIST Chemistry WebBook\footnote{https://webbook.nist.gov/chemistry} and KIDA \footnote{https://kida.astrochem-tools.org}. Detailed information can be found in Table \ref{tab:formation_enthalpies}. As little modification as possible has been made to the list of these values, as the main aim of this work is to improve the result of the S16 model. Only if necessary for the calculation of the reactive desorption efficiencies, values have been added. This concerns all values adopted from the NIST Chemistry WebBook and KIDA. For the V17 model, the enthalpies are also mostly taken from \cite{du2012} and the binding energies from \cite{minissale2016b}. Apparently, discrepancies between both codes also exist for some of the reactions involved in the hydrogenation chain towards methanol:

\begin{align}
&\text{HCO} + \text{H} \xrightarrow[\substack{\text{pyRate}: 15.4\%}]{\substack{\text{MONACO}: 5.4\%}} \text{H}_2\text{CO} \label{eq:efficiency_H2CO} \\
&\text{CH}_2\text{OH}/\text{CH}_3\text{O} + \text{H} \xrightarrow[\substack{\text{pyRate}: 0.05\%/0.08\%}]{\substack{\text{MONACO}: 0.64\%}} \text{CH}_3\text{OH} . \label{eq:efficiency_CH3OH}
\end{align}

For example for reaction \ref{eq:efficiency_H2CO} and \ref{eq:efficiency_CH3OH} the two codes yield very different reactive desorption efficiencies. In \textit{pyRate}, H$_2$CO is desorbed more efficiently before it has the opportunity to react further and eventually form CH$_3$OH, which is also less likely to desorb from the surface of the dust grain compared to the formation scenario in \textit{MONACO}.

Figure \ref{fig:comparison_V17} presents the column density profiles of CO, H$_2$CO, CH$_3$OH and CH$_2$DOH of models 1D-4, S16 and V17. In the case of V17, the CH$_2$DOH column density was derived by scaling the CH$_3$OH column density with the respective deuteration ratio from S16, as $\textit{MONACO}$ does not include a description of deuterium chemistry. Note that the column density profiles are shown at different time steps. The V17 model and the S16 model presented in \cite{chacon(2019)} showed the time steps when the CO column density in the respective model is comparable to the observed value. For V17 this corresponds to $t$ = \SI{1.6e+5}{yr}, which coincides with the peak of the COM abundances. The same estimation for S16 yields a very early time step of $t$ = \SI{3.0e+4}{yr}. The resulting column densities are shown in Figure \ref{fig:comparison_V17} as the dark red line. We have derived a new best fit time corresponding to the lowest $\chi^2$ value of the observed column densities vs. the corresponding values in the 1D-4 model, yielding t = $\SI{3.0e+5}{yr}$. This time step corresponds roughly with the peak of the methanol abundance in the 1D-4 model and is also in good agreement with the one estimated by the V17 model, varying only by a factor 2. In Figure \ref{fig:comparison_V17} we show the results for the S16 model and the 1D-4 model at the new best fit time step as bright red or green line respectively. 
The S16 model produces less non-deuterated and singly deuterated methanol at this later time step, which is caused by gas-phase chemical reactions where atom exchanges between the two functional groups of methanol were allowed. Such reactions are not allowed in the present work. The 1D-4 model, on the other hand, is more consistent with the results of the V17 model, though it still underestimates the column densities of methanol by roughly an order of magnitude. The V17 model overestimates the CH$_3$OH column density and as consequence also CH$_2$DOH column density, which was credited to a likely overestimation of the reactive desorption efficiency.
All models overestimate the amount of gas phase H$_2$CO. The V17 and S16 model, that are evaluated at a time step to match the observed CO column densities, are off by more than an order of magnitude, whereas S16 at $t$ = \SI{3.0e+05}{yr} and 1D-4 produce only twice the observed column density.
Although the V17 model and the newly developed fiducial model, 1D-4, still differ for the CH$_3$OH column density by almost two orders of magnitude (one order of magnitude of overestimation by V17 and one order of magnitude underestimation by 1D-4), we are able to confirm qualitatively some results of \cite{vasyunin2017}. The most important conclusion is that we are only able to reconcile our models with the observed values, if we consider some form of enhanced diffusion on the surface of dust grains. One possibility to get a higher diffusion rate is to enable diffusion by quantum tunneling of H and D atoms in the models. Other options are discussed below. Also, similar to \cite{vasyunin2017}, we conclude that other forms of non-thermal desorption e.g.: cosmic-ray induced desorption and photodesorption seem to have a negligible impact on the release of methanol and its precursors into the gas phase. 

\section{Discussion} \label{ch:discussion}

In addition to the comparison with the V17 model, we have explored various alterations of our model, mainly to investigate the effect on the deuteration of methanol and possibly improve the agreement between the predicted and observed $N$(CH$_2$DOH)/$N$(CH$_3$OH) ratio. An overview of the various models and their modified chemical and physical parameters is presented in Table \ref{tab:overview_models}. The 1D-4 model was picked as the fiducial model because of its closeness to V17 in terms of parameter space. However, for most of the variations of the other physical parameters, we have used the 1D-3 model as the point of reference, comprising the abstraction reactions, as most other modelling works include them. In Table \ref{tab:overview_models}, models that are derived modifying the fiducial model 1D-4 are indicated by an asterisk($^{\star}$), whilst those derived from 1D-3 are highlighted by a dagger($^{\dag }$). We have only varied one physical parameter at a time, in order to undoubtedly ascribe the altered results to the made modifications. For the discussion of the effects of the variations, we have roughly divided them into chemical and physical variations. We collected an overview of the resulting $N$(CH$_2$DOH)/$N$(CH$_3$OH) ratios for for all presented models at four different time steps for the positions of the dust peak and of the methanol peak in Table \ref{tab:D_H_ratio}.

\begin{table*}
	\caption{Overview of the various models investigated in this work.}
	\label{tab:overview_models}
	\begin{tabular}{l c c l c l l l}
	\hline
	\hline
	 $\text{Model}$ &  $\text{tunneling}$ & $\text{thermal}$ & $\text{E}_{\mathrm{d}}$/$\text{E}_{\mathrm{b}}$\tablefootmark{a}  &$\text{abstraction}$ & $\text{gas-grain}$ &  $\text{cosmic-ray}$ & $\text{other}$ \\
    &  $\text{diffusion}$ & $\text{diffusion}$ & & $\text{reactions}$ & $\text{model}$ & $\text{ionisation rate}$ & $\text{modifications}$  \\
    \hline
	\textbf{Chemical variation:} \\
	\noalign{\smallskip}
	\hline
	\noalign{\smallskip}
     $\text{1D-1}^{\dag}$ & $\text{\ding{55}}$ & $\text{\ding{51}}$ & 0.55 & $\text{\ding{51}}$ & $\text{3 phases}$ & \SI{1.3e-17}{\per\second} &  \\
    $\text{1D-2}^{\star}$ & $\text{\ding{55}}$ & $\text{\ding{51}}$ & 0.55 & $\text{\ding{55}}$ & $\text{3 phases}$ & \SI{1.3e-17}{\per\second}\\
    $\text{1D-3}$ & $\text{\ding{51}}$ & $\text{\ding{51}}$ & 0.55 & $\text{\ding{51}}$ & $\text{3 phases}$ & \SI{1.3e-17}{\per\second} & \\
    $\text{1D-4}$ & $\text{\ding{51}}$ & $\text{\ding{51}}$ & 0.55 & $\text{\ding{55}}$ & $\text{3 phases}$ & \SI{1.3e-17}{\per\second} & \\
    $\text{1D-4-1}^{\star}$ & $\text{\ding{51}}$ & $\text{\ding{51}}$ & 0.55 & $\text{\ding{55}}$ & $\text{3 phases}$ & \SI{1.3e-17}{\per\second} & $\text{4 active surface layers}$\\
    $\text{1D-4-2}^{\star}$ & $\text{\ding{51}}$ & $\text{\ding{51}}$ & 0.55 & $\text{\ding{55}}$ & $\text{2 phases}$ & \SI{1.3e-17}{\per\second} & \\
    $\text{1D-5}^{\dag}$ & $\text{\ding{55}}$ & $\text{\ding{51}}$ & 0.2 & $\text{\ding{51}}$ & $\text{3 phases}$ & \SI{1.3e-17}{\per\second} & \\
    $\text{1D-6}^{\star}$ & $\text{\ding{55}}$ & $\text{\ding{51}}$ & 0.2 & $\text{\ding{55}}$ & $\text{3 phases}$ & \SI{1.3e-17}{\per\second} & \\
    $\text{1D-7}^{\dag}$ & $\text{\ding{51}}$ & $\text{\ding{51}}$ & 0.55 & $\text{\ding{51}}$ & $\text{2 phases}$ & \SI{1.3e-17}{\per\second} & \\
    $\text{1D-8}^{\dag}$ & $\text{\ding{51}}$ & $\text{\ding{51}}$ & 0.55 & $\text{\ding{51}}$ & $\text{3 phases}$ & \SI{1.3e-17}{\per\second}  &$\text{no reactive desorption for}$ \\
    &&&&&&& $\text{reactions with 2 products}$ \\
    $\text{1D-9}^{\dag}$ & $\text{\ding{51}}$ & $\text{\ding{51}}$ & 0.55 & $\text{\ding{51}}$ & $\text{3 phases}$ & \SI{1.3e-17}{\per\second} & $\text{reduced E}_{\mathrm{A}}$\tablefootmark{b} \medspace $\text{for deuterated}$ \\
    &&&&&&& $\text{species by}$ \medspace \SI{200}{K}\\
    $\text{1D-10}^{\dag}$ & $\text{\ding{51}}$ & $\text{\ding{51}}$ & 0.55 & $\text{\ding{51}}$ & $\text{3 phases}$ & \SI{1.3e-17}{\per\second} & $\text{H}_{\mathrm{form}}$$(\text{XH})$ = $\text{H}_{\mathrm{form}}$($\text{XD}$)\tablefootmark{c} \\ 
    &&&&&&& ($\text{see Table}$ \medspace \ref{tab:formation_enthalpies})  \\
    \hline
	\noalign{\smallskip}
	\textbf{Physical variation:} \\
	\noalign{\smallskip}
	\hline
    $\text{1D-11}^{\dag}$ & $\text{\ding{51}}$ & $\text{\ding{51}}$ & 0.55 &  $\text{\ding{51}}$ & $\text{3 phases}$ & \SI{1.3e-17}{\per\second} & $\text{decreased}$ \medspace $\text{T}_{\mathrm{gas}}$ \medspace $\text{by}$ \medspace \SI{1}{K}  \\
     & & & & & & & $\text{throughout the core}$ \\
    $\text{1D-12}^{\dag}$ & $\text{\ding{51}}$ & $\text{\ding{51}}$ & 0.55  & $\text{\ding{51}}$ & $\text{3 phases}$ & $\mathscr{L}$$\text{-model}$\tablefootmark{d} & \\
    $\text{1D-13}^{\dag}$ & $\text{\ding{51}}$ & $\text{\ding{51}}$ & 0.55 & $\text{\ding{51}}$ & $\text{3 phases}$&  &$\text{dynamic mechanism}$  \\
    &&&&&&& $\text{for CR- desorption }$\tablefootmark{e} \\
    \noalign{\smallskip} 
	\hline
	\end{tabular}
	\tablefoot{Listed are only chemical and physical properties that were varied between the models.
	The 1D-4 model is the fiducial model presented in Section \ref{ch:fiducial_model}. Models taking the 1D-4 model as starting point are marked with an asterisk($^{\star}$). Models taking the 1D-3 model, including the abstraction reactions, as a starting point are marked with a dagger($^{\dag}$).
	\tablefoottext{a}{$\text{E}_{\mathrm{d}}$ \medspace $\text{is the diffusion energy and}$} \medspace $\text{E}_{\mathrm{b}}$ \medspace $\text{the binding energy}$ 
	\tablefoottext{b}{$\text{E}_{\mathrm{A}}$} \medspace $\text{is the activation energy of the reaction}$ 
	\tablefoottext{c}{$\text{H}_{\mathrm{form}}$($\text{XH}$)} \medspace $\text{is the formation enthalpy}$ \\  $\text{of a non-deuterated species and}$ \medspace $\text{H}_{\mathrm{form}}$($\text{XD}$) \medspace $\text{is the formation enthalpy of a deuterated species}$ 
	\tablefoottext{d}{$\text{from \cite{padovani2018}}$} 
	\tablefoottext{e}{$\text{from \cite{sipilae2021}}$}
	}
\end{table*}

\subsection{Chemical Variation}
\subsubsection{1D-1 to 1D-6: Combining Enhanced Diffusion and Abstraction Reactions} \label{ch:abstraction_diffusion}

As already pointed out in Section \ref{ch:comparison_V17}, to explain the observed magnitudes of molecular abundances and column densities, we have to introduce some type of enhanced diffusion of atoms on the grain surface. In V17 and the fiducial model, this is realised by enabling the diffusion of H (and D) by quantum tunneling. Laboratory experiments on an amorphous solid water surface (\citealt{hama2012}) as well as a CO surface (\citealt{kimura2018}) actually suggest that diffusion of H and D is dominated by thermal hopping even at temperatures around \SI{10}{K}. This was concluded since a significant isotope effect, which is expected in the case that the diffusion proceeds mainly via quantum tunneling, could not be observed. Therefore, we have explored both the option for an increased, fast thermal diffusion as well as the diffusion of H and D atoms by quantum tunneling. 

For the former, a value of the diffusion-to-binding energy $E_{\mathrm{d}}/E_{\mathrm{b}}$ has to be chosen. Due to the lack of detailed experimental data, most chemical codes apply only one value for the diffusion-to-binding energy to every surface species. In reality, it is likely that different species have an individual diffusion-to-binding energy, that extends over a wider range of values, depending on the various types of potential wells present. Reasonable values of the diffusion-to-binding energy cover the range of 0.2 to 0.7 (\citealt{furuya2022}). As a first approximation, we assume that the comparatively large number of hydrogen and deuterium atoms might fill up the deeper potential wells quite quickly, making the shallower potential wells more relevant for the diffusion process. For that reason, we have adopted the lowest debated value of 0.2 for the diffusion-to-binding energy.
As a reference, we have also tested the option to introduce no enhanced diffusion - neither fast thermal diffusion nor tunneling diffusion, but only to employ a typically assumed value for the diffusion-to-binding energy of 0.55 leading to slow thermal hopping.

\begin{table*}
    \centering
	\caption{$N$(CH$_2$DOH)/$N$(CH$_3$OH) ratio at the centre of the pre-stellar core $r_{\mathrm{cen}}$ and the radius of the methanol peak $r_{\mathrm{max}} = \SI{5200}{AU}$.}
	\label{tab:D_H_ratio}
	\begin{tabular}{l | l l | l l | l l | l l}
	\hline
	\hline
	\noalign{\smallskip}
	 $\text{Model}$ &  $\text{t}$ = \SI{1e5}{yr} && $\text{t}$ = \SI{3e5}{yr} && $\text{t}$ = \SI{5e5}{yr} && $\text{t}$ = \SI{1e6}{yr} \\
 	 &  $\text{r}_{\mathrm{cen}}$ & $\text{r}_{\mathrm{max}}$ & $\text{r}_{\mathrm{cen}}$& $\text{r}_{\mathrm{max}}$& $\text{r}_{\mathrm{cen}}$ & $\text{r}_{\mathrm{max}}$ & $\text{r}_{\mathrm{cen}}$ & $\text{r}_{\mathrm{max}}$ \\
 	 & [\%]  & [\%] & [\%] & [\%]  & [\%]  & [\%]  & [\%]  & [\%] \\
    \hline
	\noalign{\smallskip}
    \text{1D-1} & 3.8 & 3.3 & 5.4 & 4.6 & 6.2 & 5.0 & 6.6 & 5.1  \\
    \text{1D-2} & 3.8 & 3.3 & 5.2 & 4.5 & 5.9 & 4.8 & 6.4 & 4.9 \\
    \text{1D-3} & 3.0 & 0.5 & 3.3 & 1.1 & 2.4 & 1.6 & 4.8 & 4.2  \\
    \text{1D-4$^*$} & 8.8 & 6.2 & 8.2 & 8.0 & 7.1 & 6.9 & 6.3 & 6.1 \\
    \text{1D-4-1} & 16.1 & 4.5 & 8.0 & 7.3 & 8.1 & 7.5 & 7.7 & 7.2 \\
    \text{1D-4-2} & 2.5 & 2.3 & 3.0 & 2.5 & 3.0 & 2.2 & 2.7 & 2.0 \\
    \text{1D-5} & 1.1 & 0.2 & 1.3 & 0.5 & 1.7 & 1.3 & 2.0 & 2.3 \\
    \text{1D-6} & 5.8 & 4.9 & 8.0 & 7.7 & 11.4 & 11.3 & 13.6 & 13.5 \\
    \text{1D-7} & 0.1 & 0.04 & 0.2 & 0.04 & 0.2 & 0.04 & 0.1 & 0.03 \\
    \text{1D-8} & 4.3 & 0.8 & 6.8 & 2.3 & 5.7 & 3.0 & 3.5 & 1.4 \\
    \text{1D-9} & 7.0 & 1.4 & 2.7 & 1.1 & 3.4 & 2.2 & 6.3 & 5.5 \\
    \text{1D-10} & 3.0 & 0.5 & 3.2 & 1.1 & 2.3 & 1.5 & 4.6 & 4.0 \\
    \text{1D-11} & 3.0 & 0.5 & 4.3 & 1.4 & 2.2 & 1.4 & 4.9 & 4.2 \\
    \text{1D-12} & 0.8 & 0.1 & 2.3 & 1.1 & 3.1 & 2.2 & 7.4 & 6.9 \\
    \text{1D-13} & 3.9 & 0.7 & 2.7 & 1.0 & 2.3 & 1.5 & 4.3 & 3.7 \\
    \noalign{\smallskip} 
	\hline
	\end{tabular}
 \tablefoot{The position of the methanol peak is determined by the profile of the fiducial model at the best fit time. An overview of the chemical and physical properties of the various models is given in Table \ref{tab:overview_models}. The model marked with the asterisk is the fiducial model and $t = \SI{3e+5}{yr}$ is the best-fit time obtained as described in Section \ref{ch:fiducial_model}. The observed values are $N$(CH$_2$DOH)/$N$(CH$_3$OH)($r_{\mathrm{cen}}$) = $\num{7.1}\% \pm {1.9}\%$ for the centre and $N$(CH$_2$DOH)/$N$(CH$_3$OH)($r_{\mathrm{max}}$) = $\num{4.5}\% \pm {1.2}\%$ for a radius of \SI{5200}{AU}.}
\end{table*}

Additionally, we decided to run two models for every explored type of diffusion process: one with only addition reactions and one with addition and abstraction reactions.

Models 1D-1 and 1D-2, employing slow thermal diffusion with a diffusion-to-binding energy $E_{\mathrm{d}}$/$E_{\mathrm{b}}$ of 0.55, serve as references to the models with enhanced diffusion. Both produce CH$_3$OH column densities around 10$^7$ cm$^{-2}$, which is several orders of magnitude lower than the observed value of 10$^{13}$ cm$^{-2}$. Models 1D-3 and 1D-4, additionally employing the diffusion via tunneling for H and D atoms, as well as models 1D-5 and 1D-6, relying on fast thermal hopping, show significantly higher column densities of the order of 10$^{12}$ cm$^{-2}$. These results deviate only within a factor of 10 from the observed values and are thereby matching the observation. Therefore, we conclude that some form of enhanced diffusion of H and D atoms over the grain surface has to take place in order to reach similar column densities as the ones measured in the pre-stellar core L1544. 

The column densities in the models that employ fast thermal hopping are higher for early time steps ($t=\SI{1e+5}{yr}$) to intermediate time steps ($t=\SI{5e+5}{yr}$) - up to a factor of six for CH$_3$OH and up to a factor of four for CH$_2$DOH -, but decline earlier than in the models with tunneling diffusion. Moreover, we find that the CH$_3$OH column density profiles in the 1D-4 model including only addition reactions are lower (by up to a factor of three) than for the 1D-3 model comprising both addition and abstraction reactions. The CH$_2$DOH column density profiles, on the other hand, are higher (by up to a factor of 2.5) in the models with addition reactions only. For the 1D-5 and 1D-6 model, employing fast thermal hopping, this effect is even more pronounced with a factor of up to ten for CH$_3$OH and up to eight for CH$_2$DOH.

\begin{figure*} 
	\includegraphics*[width=1.0\textwidth]{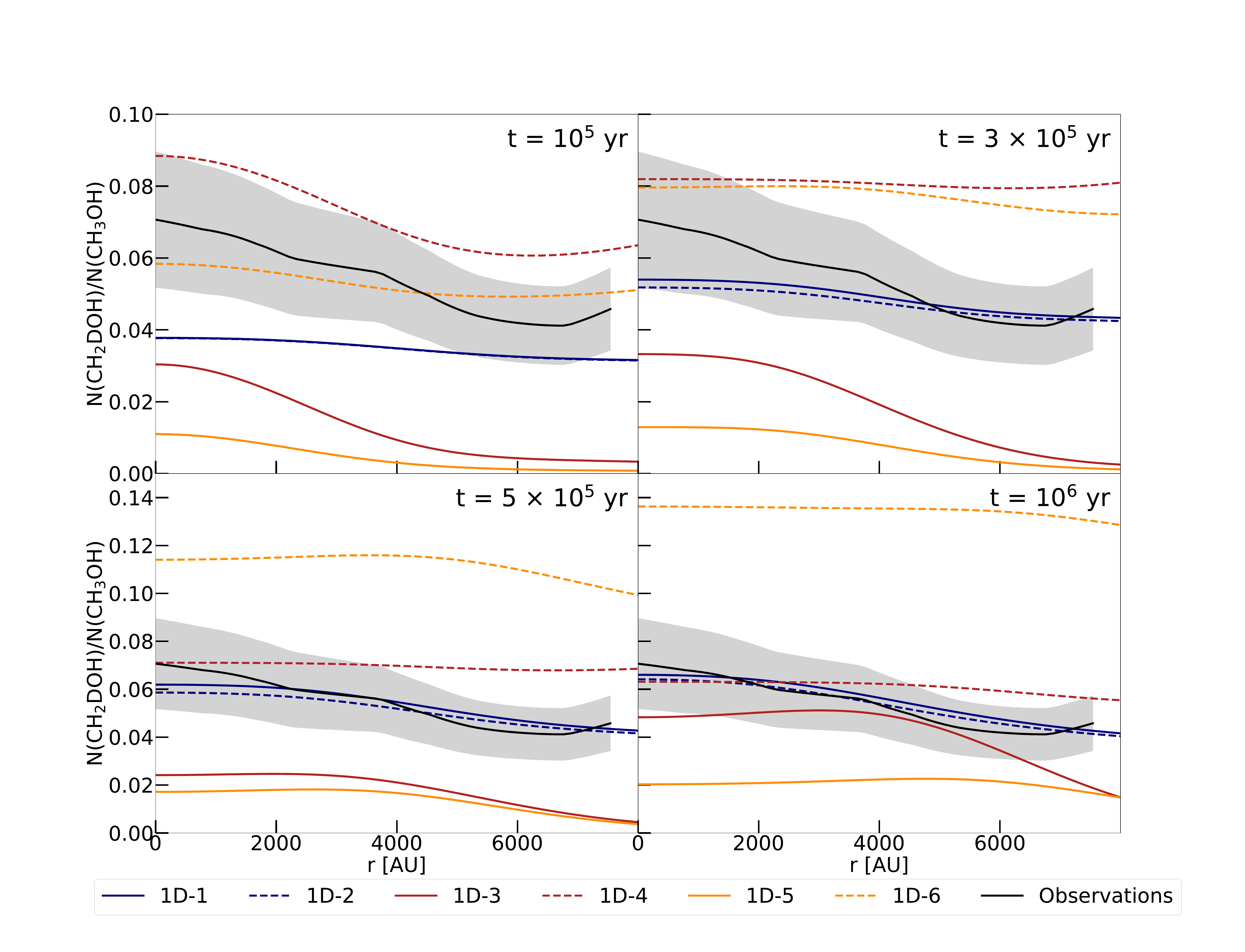}
	\caption{Modelled ratio between singly-deuterated methanol (CH$_2$DOH) and non-deuterated methanol (CH$_3$OH) for several models at four different time steps ranging from 10$^{5}$ yr to 10$^6$ yr. The best fit time is $t$ = \SI{3e+5}{yr}. We show two models with slow thermal hopping ($E_{\mathrm{d}}/E_{\mathrm{b}}$ = 0.55) (blue lines), two models with slow thermal hopping and tunneling diffusion of H and D (red lines) and two models with fast thermal hopping ($E_{\mathrm{d}}/E_{\mathrm{b}}$ = 0.2) (orange lines). The solid lines indicate models with addition and abstraction reactions, while the dashed lines indicate models with only addition reactions. The black line shows the observed ratio (errors as grey-shaded areas).}
	\label{fig:CH2DOH_CH3OH_wo_tunnel_models} 
\end{figure*}

In addition to the order of magnitude of column densities, we attempt to reproduce the deuterium fraction $N$(CH$_2$DOH)/$N$(CH$_3$OH) as closely as possible. This ratio is likely less affected by the individual modelling choices, as the effects on CH$_3$OH and CH$_2$DOH are probably similar for most parameter selections. Figure \ref{fig:CH2DOH_CH3OH_wo_tunnel_models} depicts the ratio of the column densities of singly deuterated methanol (CH$_2$DOH) and non-deuterated methanol (CH$_3$OH). Models 1D-1 and 1D-2 show a similar level of deuteration reaching a central value of $\approx \num{0.05}$ for the best fit time of $t$ = \SI{3.0e5}{yr}. 1D-1, the model that includes addition and abstraction reactions, has a slightly higher deuterium fraction than the one with only addition reactions. 
The slightly higher deuterium fraction is expected theoretically and can be explained by the following: reaction \ref{eq:form_CH2DOH} has a much higher activation energy than the competing reaction \ref{eq:form_CH3O} leading to its isomer. Therefore the hydrogenation of H$_2$CO proceeds preferably by the latter reaction:

\begin{align}
\label{eq:form_CH2DOH}
&\text{H}_2\text{CO} + \text{H} \xrightarrow{E_{\mathrm{A}} = \SI{5.16e3}{K}} \text{CH}_2\text{OH} \\
\label{eq:form_CH3O}
&\text{H}_2\text{CO} + \text{H} \xrightarrow{E_{\mathrm{A}} = \SI{2.00e3}{K}} \text{CH}_3\text{O} . 
\end{align}
However, the deuteration of CH$_3$O is not able to produce CH$_2$DOH in our chemical network, as this would require an exchange of atoms between the two functional groups. It is only able to react to CH$_3$OD in the presence of D. CH$_2$DOH can only be produced by deuteration of CH$_2$OH. Including abstraction reactions into the chemical network opens up another channel for the formation of CH$_2$DOH via the abstraction of non-deuterated methanol: 

\begin{align}
\label{eq:abstraction_CH2OH}
&\text{CH}_3\text{OH} + \text{H/D} \xrightarrow{E_{\mathrm{A}} = \num{3.62e3}/\SI{3.24e3}{K}} \text{CH}_2\text{OH} + \text{H}_2\text{/HD} \\
\label{eq:abstraction_CH3O}
&\text{CH}_3\text{OH} + \text{H/D} \xrightarrow{E_{\mathrm{A}} = \num{5.56e3}/\SI{5.16e3}{K}} \text{CH}_3\text{O} + \text{H}_2\text{/HD} .
\end{align}

Reaction \ref{eq:abstraction_CH2OH} is favoured in comparison to the analogue including its isomer \ref{eq:abstraction_CH3O}. A similar effect on the deuterium fraction of the molecular abundances was discussed for the static 0D-models in \cite{taquet2012}. They found large enhancements for the $N$(CH$_2$DOH)/$N$(CH$_3$OH) ratio in the high density case ($n_{\mathrm{H}}$ $\geq$ \SI{1e6}{\per\cubic\centi\meter}) eventually reaching values above unity, while there was no significant increase for the low density case ($n_{\mathrm{H}}$ = \SI{1e4}{\per\cubic\centi\meter} - \SI{1e5}{\per\cubic\centi\meter}). \cite{aikawa2012}, on the other hand, could not fully confirm these results with their 1D radiative hydrodynamics model. 

In contrast to the models with slow thermal diffusion, the models that include a form of enhanced diffusion - either fast thermal hopping or tunneling diffusion - show the opposite behaviour when it comes to the inclusion of abstraction reactions (see Figure \ref{fig:CH2DOH_CH3OH_wo_tunnel_models}). The models with only addition, 1D-4 and 1D-6, have an up to eight times higher $N$(CH$_2$DOH)/$N$(CH$_3$OH) ratio at certain time steps than the one where addition and abstraction reactions are included. The downside of this increased deuterium fraction is that the increased amount of CH$_2$DOH apparently causes a decrease of the amount of CH$_3$OH that is produced. 

Models 1D-3 and 1D-5, comprising addition and abstraction reactions, are not able to reproduce the observed $N$(CH$_2$DOH)/$N$(CH$_3$OH) ratio as well. They only match the level of the observations for the innermost part of the core at early time steps ($t = \SI{1e+5}{yr}$), but decline too quickly with increasing radius. For late time steps ($t=\SI{1e+6}{yr}$), the deuterium fraction is increasing again to a level, where the intermediate radii (r=2000AU - r=6000) match the observational profile. The innermost part, however, shows less deuteration than the outer ring and also the observations.

By looking at the reaction rates of for example model 1D-3, one can see that at a time step close to the peak in methanol formation ($\approx$ $t$ = \SI{5e5}{yr}), the magnitude of the addition reaction rates and the one from the abstraction reactions approach each other and basically become almost equal in value, meaning that the net formation of methanol does proceed much slower. We therefore suspect that the reaction rates for the abstraction reactions are too high/much higher than in reality.

Having assessed that some form of enhanced diffusion is needed, it is not possible to decide based on our modelling results, which of the processes of enhanced diffusion matches the observations better. For times before the time step when the methanol column density peaks, the fiducial model, employing tunneling diffusion, has a higher $N$(CH$_2$DOH)/$N$(CH$_3$OH) ratio profile than the 1D-6 model, relying on thermal hopping. The former is slightly above the observed values, but reproduces the observed shape of the profile very well. The latter is well within the area of uncertainty of the observations. It has, however, a much flatter profile than is observed. The profile of the fiducial model flattens with time, until both models become nearly identical at the best fit time of $\SI{3.0e+5}{yr}$. After the time step when the methanol column density peaks, the $N$(CH$_2$DOH)/$N$(CH$_3$OH) ratio of the fiducial model decreases and begins to match the observed values quite closely for time steps beyond $t = \SI{5.0e+5}{yr}$. The $N$(CH$_2$DOH)/$N$(CH$_3$OH) ratio of the 1D-6 model starts increasing far above the observed values, until it reaches values of almost $\num{0.14}$ for late times ($t = \SI{1e+6}{yr}$).

\subsubsection{1D-7: Variation of the Grain Model}
\label{ch:grain_model}
\begin{figure*} 
	\includegraphics*[width=1.0\textwidth]{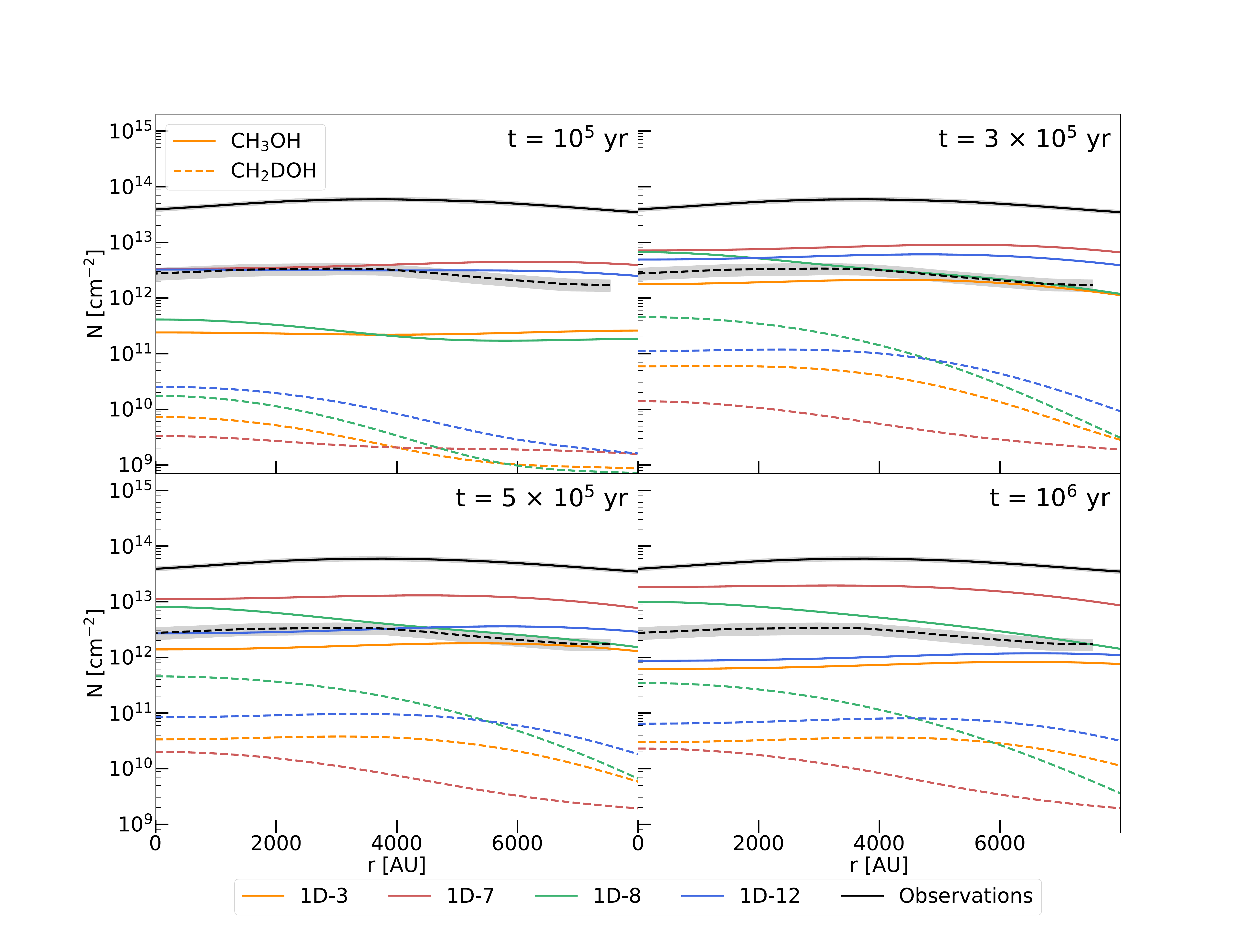}
	\caption{Modelled column density profiles of non-deuterated methanol (CH$_3$OH) and singly-deuterated methanol (CH$_2$DOH) for several models at four different
   time steps ranging from 10$^5$yr to 10$^6$yr. We present the varied models showing the largest effects on the amount of methanol in the gas phase: 1D-7 (two-phase model), 1D-8 (only reactive desorption with one product) and 1D-12 (radius-dependent cosmic-ray ionisation rate). The black lines show the observed profiles (errors as grey-shaded areas). The solid lines indicate the CH$_3$OH column densities, the dashed lines indicate the CH$_2$DOH column densities respectively.} 
	\label{fig:combined_1D7_1D8}
\end{figure*}

\begin{figure*} 
	\includegraphics*[width=1.0\textwidth]{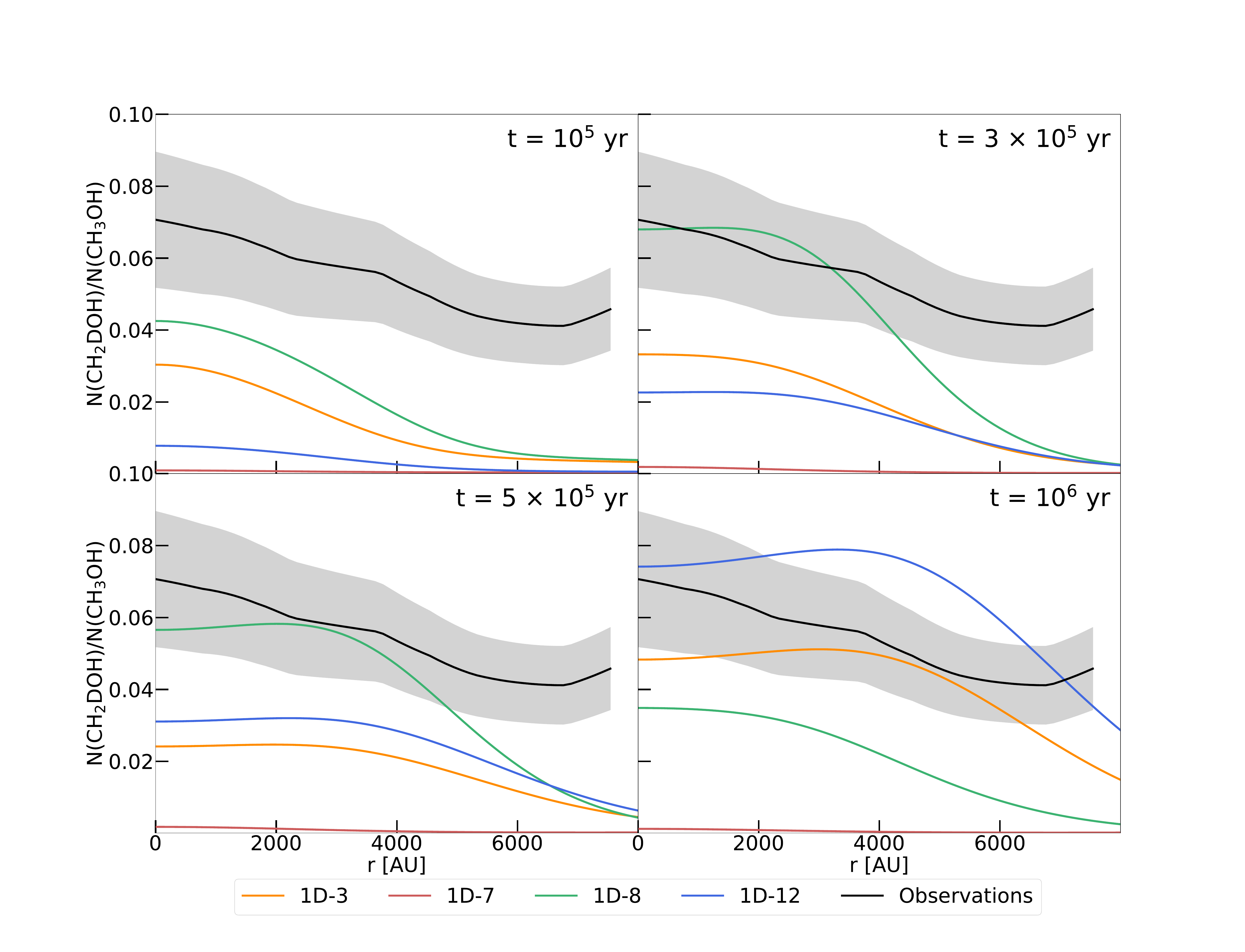}
	\caption{Modelled ratio between singly-deuterated methanol (CH$_2$DOH) and non-deuterated methanol (CH$_3$OH) for several models at four different
   time steps ranging from 10$^5$yr to 10$^6$yr. We present the varied models showing the largest effects on the amount of methanol in the gas phase: 1D-7 (two-phase model), 1D-8 (only reactive desorption with one product) and 1D-12 (radius-dependent cosmic-ray ionisation rate). The black line shows the observed ratio (errors as grey-shaded areas).}
	\label{fig:D_H_ratio_other_models}
\end{figure*}

As described in Section \ref{ch:chemical_model}, the grain model is a three-phase model, consisting of a gas phase, a chemically-active surface phase and a chemically-inert mantle phase, for most of the models presented here (see also Table \ref{tab:overview_models}). Other modelling tasks performed with $\textit{pyRate}$ (\citealt{sipilae2016a}) have shown that the choice of the grain model can have a significant effect on the magnitude of the deuterium fraction. Therefore, we decided to run a simulation with a simpler two-phase model, consisting only of a gas phase and a chemically-active surface phase. A mantle phase, where frozen-out molecules are stored without chemical alteration is not considered in this model. The 1D-7 model presented in this section is identical to the 1D-4-2 model, but for the inclusion of abstraction reactions for methanol in 1D-7 in contrast to 1D-4-2. 

The resulting column density profiles and the $N$(CH$_2$DOH)/$N$(CH$_3$OH) ratio are depicted in Figure \ref{fig:combined_1D7_1D8} and \ref{fig:D_H_ratio_other_models} respectively.
The two phase model, 1D-7, does produce a much lower deuterium fraction than the three-phase model. The deuterium fraction in the two phase model is of the order 10$^{-3}$ to 10$^{-4}$ compared to 10$^{-2}$ in the fiducial model, in model 1D-3 and in the observed $N$(CH$_2$DOH)/$N$(CH$_3$OH) ratio. This finding is consistent with previous results that deuteration on the surface is hindered by using only two phases (\citealt{sipilae2016a}).
The two-phase model produces higher column densities than model 1D-3 or model 1D-4 for CH$_3$OH. They are of the order of 10$^{13}$ cm$^{-2}$, differing only a factor of a few from the observed values. The column densities of CH$_2$DOH on the other hand are one or two orders of magnitude lower than in the three phase model and CH$_2$DOH is thereby severely under-produced.

An explanation for this behaviour is that in two phase models, deuterium atoms on the surface of dust grains tend to get locked into deuterated forms of water, ammonia and methane, forming quite stable bonds that are not easily broken up again. Additionally, the aforementioned molecules are not readily desorbed into the gas phase, which hinders the release of deuterium atoms by the dissociation of gas phase species. As a consequence, the majority of deuterium is trapped in the molecular ice contents and deuteration of other surface species, including methanol, is suppressed in two phase models compared to the more advanced multilayer models.

\subsubsection{1D-8: Variation of the Reactive Desorption Mechanism}

The reactive desorption mechanism that was in place in $\textit{pyRate}$ before the extension to the more advanced treatment, described in Section \ref{ch:reactive_desorption}, allowed only the reactive desorption of exothermic surface reactions with a single reaction product. Additionally, the mechanism applied the same reactive desorption efficiency, typically 1\%, to every eligible reaction and did not distinguish between surface types. In fact that is the case for most other reactive desorption mechanisms used in the literature, except for the one presented here and the one in \cite{vasyunin2017}. 

The new reactive desorption mechanism extends its application to chemical reactions with two reaction products. It is now interesting to quantify the consequences of this change. Note that the newly developed mechanism yields a larger reactive desorption efficiency for the lighter reaction partner and a lower efficiency for the heavier one, depending on their mass ratio. In model 1D-8, we apply the modified reactive desorption mechanism only to reactions with one product.
We anticipate that less of the light species are expelled from the surface of the dust grains. Although the reactive desorption mechanism is in place for all exothermic surface reactions, we also expect to hinder a specific effect caused by the existence of the abstraction reactions. For example for reaction \ref{eq:abstraction_reaction}:

\begin{align}
\label{eq:abstraction_reaction}
\text{H}_2\text{CO} + \text{H} \rightarrow \text{HCO} + \text{H}_2 ,
\end{align}
the new reactive desorption mechanism yields on a CO-surface an efficiency of 54\% for the H$_2$ and of \num{1.5e-36}\% for HCO. The abstraction reactions cause the expulsion of significant amounts of H$_2$, HD and D$_2$, while the desorption of the larger reaction partner is negligible.

The column densities of CH$_3$OH and CH$_2$DOH in the 1D-8 model are increased compared to the 1D-3 model (see Figure \ref{fig:combined_1D7_1D8}). The difference between the two models grows with time, beginning with a factor of $\approx$ 2 at early time steps ($t$ = \SI{1e+5}{yr}) to a factor of $\approx$ 16 at late time steps ($t$ = \SI{1e+6}{yr}) for CH$_3$OH or from $\approx$ 2.5 to $\approx$ 12 for CH$_2$DOH respectively. This results in a $N$(CH$_2$DOH)/$N$(CH$_3$OH) profile (see Figure \ref{fig:D_H_ratio_other_models}) which is very similar in shape to the 1D-3 model, but slightly higher than that of the 1D-3 model around the time steps before the methanol column density peaks in the 1D-3 model, as the CH$_2$DOH column density increases more quickly than the one of CH$_3$OH in the 1D-8 model.  This could indicate that the determined reactive desorption efficiencies do not describe the physical reality very well. In particular, that our mechanism is overestimating the efficiency with which lighter particles are expelled from the surface. However, light particles, as for example H, H$_2$ and their deuterated isotopologues, are especially important for the formation of methanol on dust grains, as it proceeds by successive addition of hydrogen and deuterium atoms. Indeed, there are hints that some of the assumptions made to set up the mechanism might not be fulfilled. For example, \cite{fredon2021} pointed out that the equal distribution of energy into all the degrees of freedom of the reaction product is unlikely to occur, and it is more likely that one or multiple degrees of freedom are favoured against the others. However, since the present work represents the first step to a more advanced treatment, we keep our assumptions simple and as general as possible. Further work could investigate different options for the partitioning of the available reaction enthalpy.

\subsubsection{1D-9: Variation of the Activation Energy}

The formation of methanol proceeds by the successive hydrogenation of CO along HCO, H$_2$CO and  CH$_2$OH/CH$_3$O to CH$_3$OH. Reactions \ref{eq:CO_HCO} to
\ref{eq:H2CO_CH3O}:

\begin{align} 
\label{eq:CO_HCO}
\text{CO} + \text{H} &\xrightarrow{E_{\mathrm{A}} = \SI{1.76e3}{K}} \text{HCO} \\
\label{eq:H2CO_CH2OH}
\text{H}_2\text{CO} + \text{H} &\xrightarrow{E_{\mathrm{A}} = \SI{5.16e3}{K}} \text{CH}_2\text{OH} \\
\label{eq:H2CO_CH3O}
\text{H}_2\text{CO} + \text{H} &\xrightarrow{E_{\mathrm{A}} = \SI{2.00e3}{K}} \text{CH}_3\text{O}
\end{align}
have an activation barrier. Their corresponding activation energies $E_{\mathrm{A}}$ are indicated on top of the arrows. The remaining reactions are barrierless. The reactions leading to deuterated analogues of these species have similar activation energies, at times with somewhat lower values. A complete overview is shown in Appendix \ref{ch:App_B} for both addition and abstraction reactions.  

The exact values of the activation energies, especially the difference for reactions leading to non-deuterated and deuterated isotopologues, could potentially have a large impact on the $N$(CH$_2$DOH)/$N$(CH$_3$OH) ratio. Therefore we explore how the results are affected by a small variation of the activation energy. Specifically, we aim to test if decreasing the activation energies for reactions producing deuterated isotopologues could lead to a significant increase of the $N$(CH$_2$DOH)/$N$(CH$_3$OH) ratio. Hence, we decided to decrease the activation energy for those reactions by \SI{200}{K}. We only varied the activation energy of specific addition reactions (marked with an asterisk in the overview in \ref{ch:App_B}). The abstraction reactions were left untouched.

Undertaking this variation produces a slightly higher CH$_2$DOH column density (see Figure \ref{fig:cd_unimportant_models}). For early times ($t$=\SI{1e+5}{yr}), the 1D-9 model shows a twice as high CH$_2$DOH column density as compared to the 1D-3 model. The difference between the two models decreases towards the time step when the methanol column density peaks at $t$ = \SI{3.0e+5}{yr}. At this time step, the 1D-8 model has a slightly lower CH$_2$DOH column density profile as compared to the 1D-3 model. After the temporal methanol peak, the difference increases for intermediate ($t$ = \SI{5.0e+5}{yr}) and late ($t$ = \SI{1e6}{yr}) time steps to a factor between 1 and 2. The CH$_3$OH column densities differ by a negligible amount between the two models (see Figure \ref{fig:cd_unimportant_models}). Consequently, the $N$(CH$_2$DOH)/$N$(CH$_3$OH) profile (see Figure \ref{fig:DH_unimportant_models}) experiences an increase as well. For early time steps the $N$(CH$_2$DOH)/$N$(CH$_3$OH) ratio reaches high values of up to 16.6 at t=$\SI{1.15e+5}{yr}$ and is for several other time steps well within the area of uncertainty of the observed profile for the very inner centre of the core (\num{0.055} - \num{0.091}) but is declining significantly more steeply at larger radii. The ratio decreases rapidly at the time when the methanol column density peaks, and increases again at late times (t=\SI{1e+6}{yr}), hitting the area of uncertainty, but presenting only a small upward shift compared to the 1D-3 model.

\subsubsection{1D-10: Variation of the Formation Enthalpies}

The incorporation of the more sophisticated reactive desorption mechanism required to expand the list of tabulated formation enthalpies $H_{\mathrm{form}}$, which are necessary to compute the reaction enthalpies $\Delta H$. The complete list of formation enthalpies (and binding energies) of species involved in exothermic surface reactions is shown in Appendix \ref{ch:App_A}. Since experimental data for deuterated isotopologues is scarce, we have for the most part adopted the same formation enthalpy values for the deuterated isotopologues as for their non-deuterated counterparts. In a few cases, however, we were able to find experimentally measured values for the deuterated analogues in the NIST Chemistry WebBook \footnote{https://webbook.nist.gov/chemistry/}. The adopted values are marked with a star in table \ref{tab:formation_enthalpies}. The formation enthalpies of the non-deuterated molecules do not differ strongly from
the values of their deuterated analogues. For most species, there is a difference of approximately $\SI{4}{kJ/mol}$ or less for singly deuterated, up to $\SI{7}{kJ/mol}$ for doubly deuterated and up to $\SI{13}{kJ/mol}$ for triply deuterated analogues.

This list of formation enthalpies and binding energies was used for all the presented models. In order to secure that changing the formation enthalpies for only some of the deuterated isotopologues has no significant effect, we have a run a model where we adopt the same values for non-deuterated and deuterated isotopologues. The column density profiles for CH$_3$OH and CH$_2$DOH of this model are shown in Figure \ref{fig:cd_unimportant_models} and the deuterium fraction profiles are shown in Figure \ref{fig:DH_unimportant_models}. The effect on both the CH$_3$OH and CH$_2$DOH column densities and on the $N$(CH$_2$DOH)/$N$(CH$_3$OH) ratio is vanishingly small. The difference with respect to the reference model, 1D-3, ranges around 4\% for the best fit time ($t$ = \SI{3.0e+5}{yr}.)

\subsection{Physical variation}

\subsubsection{1D-11: Variation of the Gas Temperature}

Based on observations of the NH$_3$ (1,1) and (2,2) lines and especially their relative strengths, there is reason to believe that the determined gas temperatures in the used physical model by \cite{keto&caselli2010} are too high at the intermediate densities, where the maximum of the CO desorption and methanol formation occurs. In principle, lower temperatures should help to promote the deuteration process. Therefore, we decided to test a model, in which we decreased the gas temperature throughout the entire core by $\SI{1}{K}$, as a first approximation for a revised temperature profile. 

The obtained column density profiles for both isotopologues of methanol, non-deuterated and singly deuterated (see Figure \ref{fig:cd_unimportant_models}), are quite close to the reference model 1D-3. The CH$_3$OH column densities of the 1D-3 model are a bit higher until the time step when the methanol column density peaks, which is reversed after the methanol column densities start decreasing again. The CH$_2$DOH column densities in the 1D-11 model are a little higher than in the 1D-3 model around the methanol peak, but are lower before and almost identical after the peak. The deuterium fraction in the 1D-11 model (see Figure \ref{fig:DH_unimportant_models}) increases less quickly, but the maximum value does reach a slightly higher maximum value than the 1D-3 model at a later time step. After the time step of the methanol peak the $N$(CH$_2$DOH)/$N$(CH$_3$OH) ratios become very similar.

\subsubsection{1D-12: Variation of the Cosmic-Ray Ionisation Rate}

UV photons are not able to penetrate the inner, denser parts of molecular clouds with visual extinctions $A_{\mathrm{V}}$ $\geq$ 1, as they are already efficiently absorbed by the outer layers of the cloud. Therefore, cosmic rays take their place as the main ionising agents in the central parts, constituting the start of the ion-molecule chemistry. The penetrating cosmic rays will ionise molecular hydrogen to form H$_2^+$, which then in turn reacts again with the large reservoir of hydrogen molecules, thereby forming the H$_3^+$ ion. This particular ion can react with deuterated molecular hydrogen HD in the following reversible reaction: 

\begin{align}
\text{H}_3^+ + \text{HD} \rightleftharpoons \text{H}_2\text{D}^+ + \text{H}_2 .
\end{align}
The direction of the reaction from left to right is strongly favoured, due to the lower zero-point energy of H$_2$D$^+$, for temperatures below $\SI{30}{K}$, that pre-stellar cores usually exhibit (strictly true only if all reactants and products are in para form \citealt{pagani1992}). Additionally, CO, the main destroyer of H$_3^+$ is mostly frozen out on dust grains in the very inner part of the pre-stellar core. These convenient conditions promote a very efficient deuteration process in the inner parts of the core, which also quickly translates the high D/H ratios to more complex molecules.

While we have used the canonical value of the cosmic-ray ionisation rate per hydrogen molecule $\zeta$(H) = \SI{1.3e-17}{\per\second} for most of the models presented here, for the 1D-12 model we have used a physical model that is attenuating the cosmic-ray ionisation rate depending on its distance from the centre. The $\mathscr{L}$-model, presented in \cite{padovani2018} and already tested in the context of the pre-stellar core L1544 by \cite{redaelli2021}, increases the cosmic-ray ionisation rate from $\zeta$(H) = \SI{2.02e-17}{\per\second} in the innermost cell to $\zeta$(H) = \SI{4.89e-17}{\per\second} at the outer boundary of the core.

The 1D-12 model exhibits higher CH$_3$OH and CH$_2$DOH column densities as compared to the 1D-3 model (see Figure \ref{fig:combined_1D7_1D8}). Especially for early times ($t$ = \SI{1e5}{yr}), the CH$_3$OH column densities are a factor of 13 higher than in the 1D-3 model. However, the difference to the 1D-3 model is decreasing over time: at intermediate times ($t$ = \SI{5e+5}{yr}) it is approximately a factor of two and even lower at late times ($t$ = \SI{1e6}{yr}). The shape of the column density profile (see Figure \ref{fig:D_H_ratio_other_models}) is almost identical between the two models, with small deviations at early times. 
For the increase of the singly deuterated methanol, we see a time-delayed behaviour compared to the non-deuterated isotopologue. CH$_2$DOH is amplified by a factor of three at early times ($t$ = \SI{1e5}{yr}), but it never gets to the values observed for CH$_3$OH. Nevertheless, the amplification for CH$_2$DOH overtakes the one for CH$_3$OH after the time step when the methanol column density peaks, which results in a larger deuterium fraction magnitude than in the 1D-3 model for the later time steps. The largest $N$(CH$_2$DOH)/$N$(CH$_3$OH) ratio of $\approx 0.74$ is reached at late times ($t$ = \SI{1e6}{yr}).

\subsubsection{1D-13: Variation of the Cosmic-Ray Desorption Mechanism}
A frequently adopted model for the cosmic-ray induced desorption (CRD) is the one laid out in \cite{hasegawa1993}. It is also used for all the models presented in this work so far. The model assumes that cosmic rays in the \num{20}-\num{70}MeV nucleon$^{-1}$ energy range deposit \SI{0.4}{MeV} of energy into the dust grain, heating it up to a temperature $T_{\mathrm{max}}$ of \SI{70}{K}.
The CRD rate coefficient for molecule $i$ is calculated as the product of the thermal desorption rate $k_{\mathrm{max}}$(i, T$_{\mathrm{max}}$) of $i$ at the temperature $T_{\mathrm{max}}$ and an efficiency term $f(a, T_{\mathrm{max}})$ for a grain of radius $a$: 
\begin{align}
    k_{\mathrm{CR}}(i) = f(a, T_{\mathrm{max}}) k_{\mathrm{therm}}(i, T_{\mathrm{max}}) .
\end{align}
The efficiency factor $f(a, T_{\mathrm{max}})$ is determined as the ratio between the cooling time of the grains $\tau_{\mathrm{cool}}$ to the heating interval $\tau_{\mathrm{heat}}$. The \cite{hasegawa1993} model adopts constant values for these quantities e.g.: $f(a, T_{\mathrm{max}})$ = \SI{1e-5}{s}/\SI{3.16e13}{s} = \num{3.16e-19} for a grain of \SI{0.1}{\mu m}. A revised version of  CRD presented by \cite{sipilae2021} refines the description of the process by introducing two major modifications to the established scheme. On one hand, the grain cooling time $\tau_{\mathrm{cool}}$ is determined now by a dynamic mechanism taking into account the individual sublimation rates of the surface molecules as a function of their time-dependent ice abundances. On the other hand, several different CR fluxes can be considered for the calculation of the heating intervals $\tau_{\mathrm{heat}}$.

In order to test how this new mechanism affects the formation of methanol and its deuterated isotopologues, we have chosen the CR flux presented in \cite{leger1985}, as this is the one most consistent with the canonical value of the cosmic-ray ionisation rate $\zeta$(H) = \SI{1.3e-17}{\per\second}, which we used for the other models. The impact on singly and non-deuterated methanol formation seems to be minor compared to the 1D-3 model. This results fits well with the finding that the desorption of species involved in the methanol formation scheme is dominated by reactive desorption, rather than cosmic-ray induced desorption.
The CH$_3$OH column density profile (see Figure \ref{fig:cd_unimportant_models}) is slightly decreased, especially for the earlier time steps , while the CH$_2$DOH profile is not significantly changed, resulting in a little higher $N$(CH$_2$DOH)/$N$(CH$_3$OH) ratio (see Figure \ref{fig:DH_unimportant_models}). It reaches its highest value of $\num{0.088}$ at $t$ = \SI{1e+5}{yr} in the very inner centre. However, the decline is again much steeper than is observed.

\section{Conclusion} \label{ch:conclusion}

We presented several models for the prediction of column densities and deuterium fractions of methanol and its deuterated isotopologues in pre-stellar cores. As a comparison to observed quantities, we use single-dish observations of H$_2$CO and CH$_3$OH and some of their deuterated isotopologues towards the pre-stellar core L1544 conducted and analysed by \cite{chacon(2019)}.

All introduced models use a novel treatment of reactive desorption of molecules from the surface of interstellar dust grains. The treatment is based on experimental justification (\citealt{minissale2016b}) and derives an individual reactive desorption efficiency for every species, depending on the forming chemical reaction(s), and on the type of underlying surface. 

Our fiducial model serves as a comparison to the results of the models V17 (\textit{MONACO}) and S16 (\textit{pyRate}), presented in \cite{chacon(2019)}. It includes thermal diffusion (diffusion-to-binding energy ratio $E_{\mathrm{d}}/E_{\mathrm{b}}$ = 0.55) as well as the diffusion of hydrogen and deuterium atoms by quantum tunneling. The chemical network does not comprise abstraction reactions for the methanol reaction scheme. We estimate a best fit time, which coincides with the occurrence of the methanol peak at \SI{3.0e+5}{yr}.
At this time step, we find a better agreement with the observations than for the S16 model. The column densities of CH$_3$OH and CH$_2$DOH are still underestimated. However, instead of more than two orders of magnitude deviation, we are able to reduce the difference to approximately an order of magnitude. This improvement is not possible without increasing the diffusion rate on the surface of the dust grain. The observed $N$(CH$_2$DOH)/$N$(CH$_3$OH) can be reproduced quite closely. Additionally, we find that increasing the number of layers in the chemically active surface phase from one to four, which allows the atoms to also diffuse in vertical direction and as is done in the V17 model, has an increasing effect on the column densities of non-deuterated and singly deuterated methanol as well as their D/H ratio in the time frame in question.  

Previous work by \cite{vasyunin2017} and \cite{chacon(2019)} showed on the one hand the necessity to employ an increased rate of surface diffusion and on the other hand to disregard abstraction reactions from the reaction scheme, in order to reproduce the observed order of magnitude for the methanol abundances. Therefore, we have also explored various types of diffusion processes: slow thermal hopping ($E_{\mathrm{d}}/E_{\mathrm{b}}$ = 0.55), fast thermal hopping ($E_{\mathrm{d}}/E_{\mathrm{b}}$ = 0.2) and slow thermal hopping ($E_{\mathrm{d}}/E_{\mathrm{b}}$ = 0.55) combined with the diffusion of H and D atoms by quantum tunneling. 
From these tests, we conclude that a form of enhanced diffusion over the surface needs to take place to explain the observational results. Only employing slow thermal hopping produces CH$_3$OH and CH$_2$DOH column densities several orders of magnitudes below the observed ones. However, we can not decide based solely on our models which enhanced diffusion process --- fast thermal hopping or tunneling diffusion --- matches the observed $N$(CH$_2$DOH)/$N$(CH$_3$OH) ratio profiles better. Both produce D/H ratios of a similar level, which are in good agreement with the observed values. For the best fit time, the two options have nearly identical profiles.

Also, we have tested both options, employing either addition and abstraction reactions or only addition reactions, for every explored diffusion process. In general, we conclude that including abstraction reactions following \cite{hidaka2009}, in combination with an increased diffusion rate, leads to $N$(CH$_2$DOH)/$N$(CH$_3$OH) ratios that are a factor of a few lower than in the models only including addition reactions. We ascribe this behaviour to the fact that the reaction rates of the abstraction reactions become comparable to the addition reactions when combined with enhanced diffusion processes.

Furthermore, we have explored other modifications to our model, that we were suspecting to have an effect on the $N$(CH$_2$DOH)/$N$(CH$_3$OH) ratio:

\begin{itemize}
    \item \textbf{two-phase model:} higher CH$_3$OH and CH$_2$DOH column densities; deuterium fraction is severely underestimated by more than one order of magnitude
    \item \textbf{reactive desorption applied only to reactions with one reaction product:} higher CH$_3$OH and CH$_2$DOH column densities, resulting in a slightly better agreement with the observations 
    \item \textbf{location-dependent cosmic-ray ionisation rate:} higher CH$_3$OH and CH$_2$DOH column densities only differing with the observations by a factor of 2 to 3.
\end{itemize}
Only small effects on the CH$_3$OH and CH$_2$DOH column density and deuterium fraction profiles were found for the following models:
\begin{itemize}
    \item decrease of the activation energy by \SI{200}{K} leading to deuterated isotopologues compared to non-deuterated species
    \item inclusion of individual formation enthalpies for some deuterated isotopologues as opposed to using the same formation enthalpies for hydrogenated and deuterated isotopologues
    \item decrease of the gas temperature by \SI{1}{K} throughout the entire core
    \item refinement of the used cosmic-ray desorption mechanism following \cite{sipilae2021}.
\end{itemize}

Further work needs to be carried out on quantifying the reactive desorption mechanism. On the one hand, there is reason to question the assumption of equal partitioning of energy into all the degrees of freedom. Also, some of our results could hint to the fact that light particles are desorbed too easily into the gas phase with the employed mechanism. On the other hand, it would be interesting to investigate how the reactive desorption mechanism influences other species that are formed on the surface of dust grains.

Additionally, a closer look into the intricacies of the surface diffusion processes, especially of H and D, is needed. Particularly interesting would be to explore the effects of introducing different types of potential wells in which species can be trapped. Other reaction mechanism as the Eley-Rideal mechanism, which is not treated by many chemical codes, or non-diffusive chemistry could play an important role for the formation and deuteration of methanol. These mechanisms will be the subject of a future paper.

We conclude that to obtain a reasonable match with the observational column density and deuterium fraction profiles, it is necessary to employ a form of enhanced diffusion process - either fast thermal hopping or diffusion by quantum tunneling. Furthermore, the inclusion of abstraction reactions in the methanol formation scheme, while also using a fast diffusion process leads to $N$(CH$_2$DOH)/$N$(CH$_3$OH) ratios that are a factor of a few lower than without the abstraction reactions.

\begin{appendix}
\onecolumn
\section{Table of Formation Enthalpies and Binding Energies} \label{ch:App_A}
In table \ref{tab:formation_enthalpies}, we present the formation enthalpies H$_{\mathrm{form}}$ and binding energies E$_{\mathrm {b}}$ for all species involved in exothermic surface reactions, on which we applied the newly developed reactive desorption mechanism laid out in Section \ref{ch:reactive_desorption}.

\begin{table}[h!]
	\caption{Tabulated formation enthalpies H$_{\mathrm{form}}$ and binding energies E$_{\mathrm {b}}$.} 
	\label{tab:formation_enthalpies}
	\begin{tabular}{l l l | l l l}
	\hline
	\hline
	\noalign{\smallskip}
	 $\text{Species}$ & $\text{H}_{\mathrm{form}}$ & $\text{E}_{\mathrm{b}}$ & $\text{Species}$ & $\text{H}_{\mathrm{form}}$ & $\text{E}_{\mathrm{b}}$ \\
	  &  [$\text{kJ mol}^{-1}$] & [$\text{K}$] &&  [$\text{kJ mol}^{-1}$] & [$\text{K}$]\\
	\noalign{\smallskip}
	\hline
	\noalign{\smallskip}
	C & 716.70 & 800.0 & O$_2$ & 0.00 & 1000.0 \\
	CH & 594.10 & 925.0 & O$_2$H & 2.10 & 3650.0 \\
	CH$_2$ & 386.40 & 1050.0 & O$_3$ & 142.70 & 1800.0 \\ 
	CH$_3$ & 145.70 & 1175.0 & OCN & 159.40 & 2400.0 \\
	CH$_3$O & 17.00 & 3800.0 & OCS & 138.40 & 2888.0 \\
	CH$_2$OH & -9.00 & 5084.0 & OH & 39.00 & 2850.0 \\
	CH$_3$OH & -201.20 & 5534.0 & OD & 36.60$^{a}$ & 2850.0 \\
	CH$_4$ & -74.90 & 1300.0 & S & 277.00 & 1100.0 \\
	CN & 435.10 & 1600.0 & SO & 5.00 & 2600.0\\
	CO & -110.50 & 1150.0 & SO$_2$ & -296.80 & 3405.0\\
	CO$_2$ & -393.50 & 2575.0 & C$_2$ & 837.74 & 1600.0 \\
	CS & 280.30 & 1900.0 & CCH & 476.98 & 2137.0  \\ 
	H & 218.00 & 450.0 & C$_2$H$_2$ & 226.73 & 2587.0 \\
	D$^{\star}$ & 221.72$^{a}$ & 450.0 & C$_3$ & 820.06$^{a}$ & 2400.0 \\
	H$_2$ & 0.0 & 500.0 & C$_4$ & 970.69$^{a}$ & 3200.0 \\
	HD$^{\star}$ & 0.32$^{a}$ & 500.0 & C$_5$ & 979.06$^{a}$ & 4000.0 \\
	H$_2$CO & -115.90 & 2050.0 & C$_2$H$_3$ & 299.00$^{a}$ & 3037.0 \\
	H$_2$O & -241.80 & 5700.0 & C$_4$H$_2$ & 464.00$^{a}$ & 4187.0 \\
	HDO$^{\star}$ & -245.37$^{a}$ & 5700.0 & C$_2$H$_4$ & 52.40$^{a}$ & 3487.0 \\
	D$_2$O$^{\star}$ & -249.20$^{a}$ & 5700.0 & C$_2$H$_5$ & 119.00$^{a}$ & 3937.0  \\
	H$_2$O$_2$ & -135.80 & 5700.0 & HC$_3$N & 354.00$^{a}$ & 4580.0 \\
	H$_2$S & -20.5 & 2743.0 & H$_2$CS & 118.00$^{a}$ & 2700.0 \\
	D$_2$S$^{\star}$ & -23.89$^{a}$ & 2743.0 & MgH & 169.03$^{a}$ & 5750.0 \\
	HCN & 135.10 & 2050.0 & NaH & 124.27$^{a}$ & 12250.0 \\
	HNC & 135.10 & 2050.0 & PH & 253.55$^{a}$ & 5000.0 \\
	HCO & 43.50 & 1600.0 & PH$_2$ & 125.94$^{a}$ & 5000.0 \\
	HCS & 296.20 & 2350.0 & PH$_2$ & 125.94$^{a}$ & 5000.0 \\
	HNO & 99.60 & 2050.0 & SiH & 376.66$^{a}$ & 3150.0 \\
	HS & 139.30 & 1450.0 & SiH$_4$ & 34.31$^{a}$ & 4500.0 \\
	DS$^{\star}$ & 138.49$^{a}$ & 1450.0 & NS & 263.59$^{a}$ & 1900.0 \\
	N & 472.70 & 800.0 & CCO & 286.60$^{a}$ & 1950.0 \\
	N$_2$ & 0.00 & 1000.0 & C$_4$H & 775.02$^{b}$ & 3737.0 \\
	N$_2$H & 245.20 & 1450.0 & C$_6$ & 1261.02$^{b}$ & 4800.0  \\ 
	N$_2$H$_2$ & 213.00 & 4756.0 & C$_6$H & 991.80$^{b}$ & 5337.0 \\
	N$_2$D$_2$ & 207.11$^{a}$ & 4756.0 & C$_7$ & 1309.34$^{b}$ & 5600.0 \\
	NH & 376.60 & 2378.0 &	l-C$_3$H & 714.09$^{b}$ & 2937.0 \\
	ND$^{\star}$ & 375.31$^{a}$ & 2378.0 & c-C$_3$H & 714.09$^{b}$ & 2937.0 \\
	NH$_2$ & 190.40 & 3956.0 & C$_5$H$_2$ & 690.36$^{b}$ & 4987.0\\
	ND$_2^{\star}$ & 185.35$^{a}$ & 3956.0 & H$_2$CN & 242.23$^{b}$ & 2400.0 \\
	NH$_2$CHO & -186.00 & 5556.0 & l-C$_3$H$_2$ & 650.36$^{b}$ & 3387.0\\
	NH$_3$ & -45.90 & 5534.0 & c-C$_3$H$_2$ & 477.96$^{b}$ & 3387.0 \\
	ND$_3^{\star}$ & -58.58$^{a}$ & 5534.0 & C$_4$H$_3$ & 545.65$^{b}$ & 4637.0 \\ 
	NO & 90.30 & 1600.0 & SiH$_2$ & 275.00$^{b}$ & 3600.0 \\
	NO$_2$ & 33.10 & 2400.0 & SiH$_3$ & 204.09$^{b}$ & 4050.0\\
    O & 249.20 & 1390.0 & HOOH & -129.89$^{b}$ & 5700.0 \\ 	
	\noalign{\smallskip} 
	\hline
	\end{tabular}
	\tablefoot{Species marked with a star $^{\star}$ are newly added formation enthalpies for deuterated isotopologues. If not stated otherwise the formation enthalpies are adopted from \cite{du2012}. The values marked with $^a$ are adopted from the NIST Chemistry WebBook \tablefootmark{1} and the ones marked with $^{b}$ are from the Kinetic Database for Astrochemistry \tablefootmark{2}. The binding energies are taken from \cite{semenov2010}.
	\tablefoottext{1}{https://webbook.nist.gov/chemistry} 
	\tablefoottext{2}{https://kida.astrochem-tools.org}
	}
\end{table}

\clearpage
\section{Applied Reaction Scheme for the Formation of Methanol and Deuterated Isotopologues} \label{ch:App_B}
\begin{figure*}[h!] 
	\includegraphics*[width=0.9\textwidth]{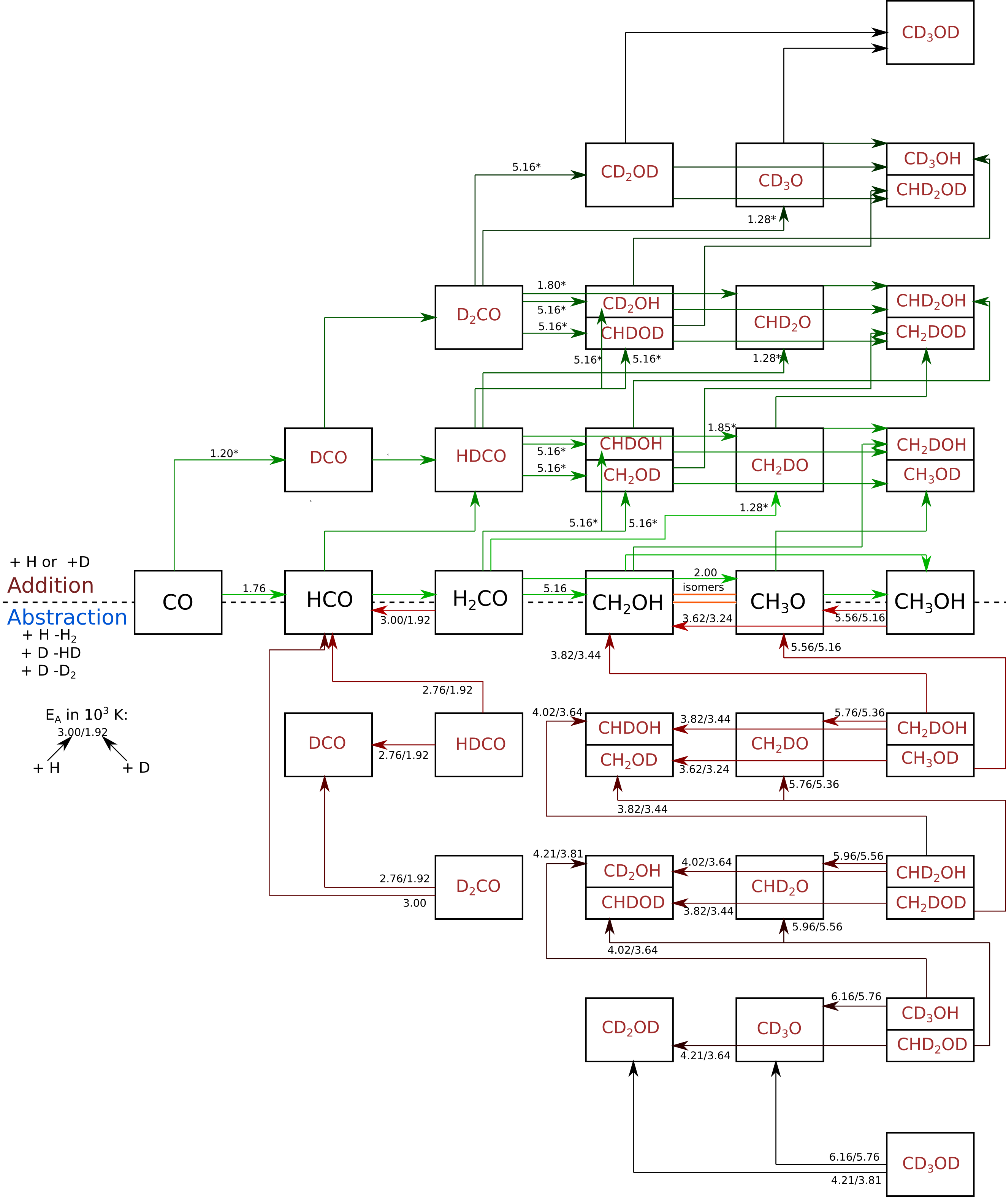}
	\caption{Reaction scheme for the formation of CH$_3$OH and its deuterated isotopologues. The upper half shows the applied "forward reactions"/addition reactions, either adding H (in horizontal direction) or D (in vertical direction). The lower half shows the employed "backward reactions"/abstraction reactions, reacting with H or D and thereby removing a H$_2$, HD or D$_2$ molecule. The values on top of the arrows indicate, if existing, the activation energies E$_{\mathrm{A}}$ in 10$^3$ K. There are always two values for the abstraction reactions. The first one gives the value for the reaction with H, the latter one for the reaction with D. Note that our reaction scheme also includes two "substitution reactions" exchanging one D atom for a H atom in H$_2$CO and HDCO respectively, which are not shown here for the sake of clarity. }
	\label{fig:network_EA} 
\end{figure*}

Figure \ref{fig:network_EA} depicts the reaction scheme for the formation of methanol, that we employed for all the presented models.

\section{Column Density and Deuterium Fraction Profiles}
\begin{figure*}[h!] 
	\includegraphics*[width=1.0\textwidth]{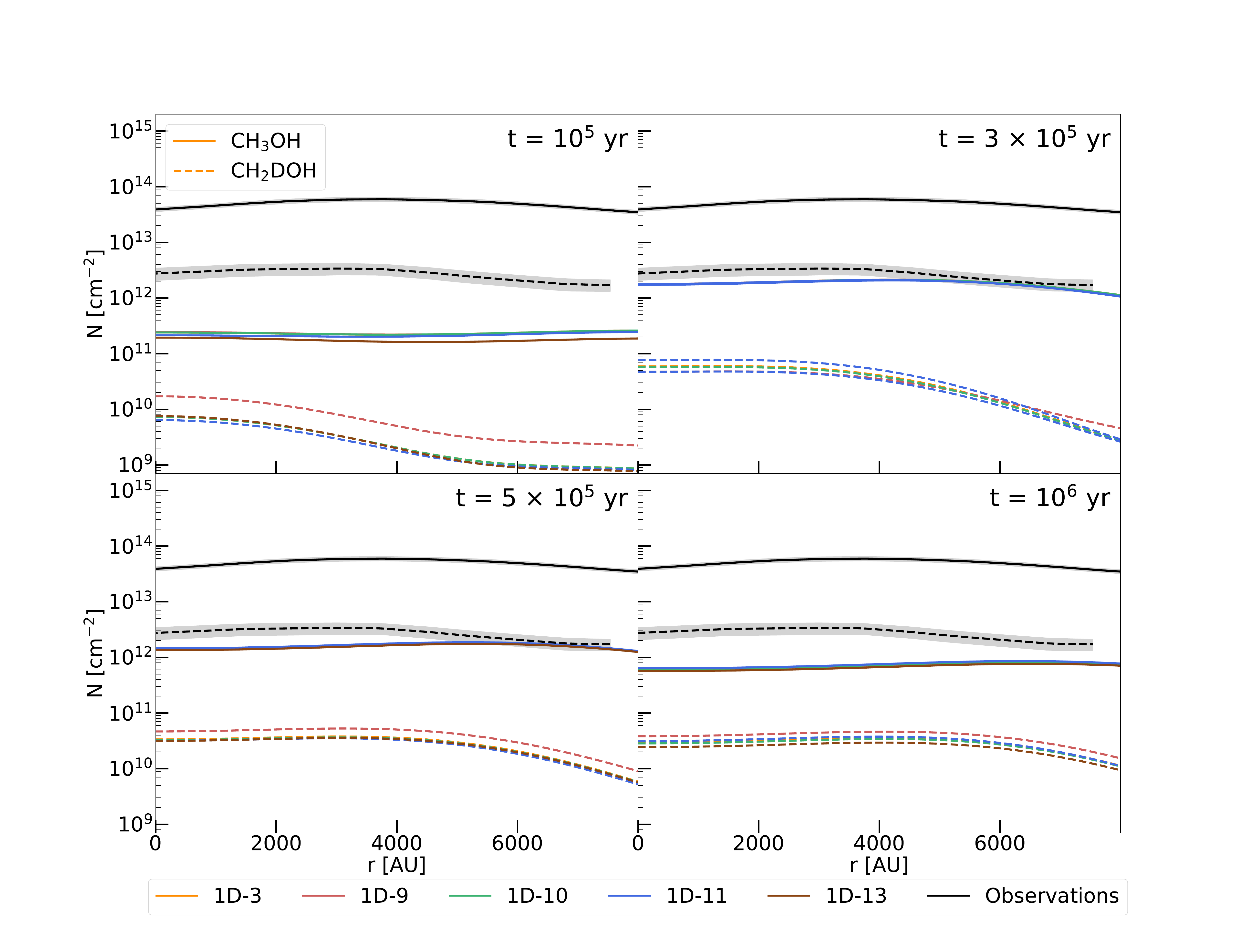}
	\caption{Modelled column density profiles of non-deuterated methanol (CH$_3$OH) and singly-deuterated methanol (CH$_2$DOH) for several models at four different
   time steps ranging from 10$^5$yr to 10$^6$yr. The black lines show the observed profiles (errors as grey-shaded areas). The solid lines indicate the CH$_3$OH column densities, the dashed lines indicate the CH$_2$DOH column densities respectively.}
	\label{fig:cd_unimportant_models} 
\end{figure*}

\begin{figure*}[h!] 
	\includegraphics*[width=1.0\textwidth]{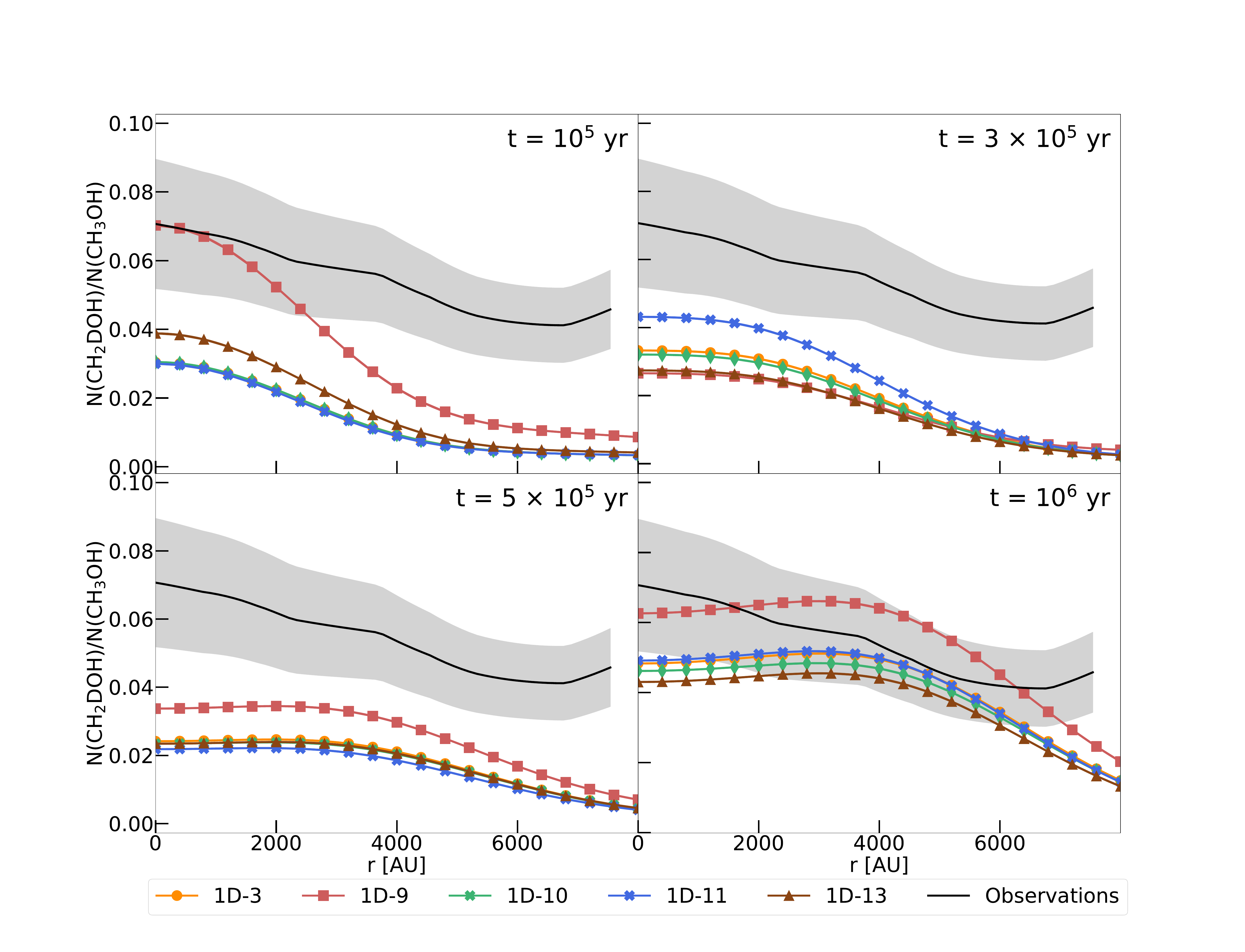}
	\caption{Modelled ratio between singly-deuterated methanol (CH$_2$DOH) and non-deuterated methanol (CH$_3$OH) for several models at four different time steps ranging from 10$^5$yr to 10$^6$yr. The black line shows the observed ratio (errors as grey-shaded areas).}
	\label{fig:DH_unimportant_models} 
\end{figure*}

Figures \ref{fig:cd_unimportant_models} and \ref{fig:DH_unimportant_models}, respectively show the modelled column density and deuterium fraction profiles of the models with only small effects.  We present the models 1D-9 (decreased activation energy for deuterated isotopologues), 1D-10 (individual formation enthalpies for deuterated isotopologues), 1D-11 (decrease of the gas temperature) and 1D-13 (refinement of cosmic-ray desorption mechanism).

\end{appendix}
\end{document}